\documentclass[useAMS,usenatbib]{mn2e}
\usepackage{graphicx}
\usepackage{grffile}\usepackage{epsf}
\usepackage{amsmath}
\usepackage{float}
\usepackage{fixltx2e}

\newcommand{\kmsmpc}{\kms\;{\rm Mpc}^{-1}}

\newcommand{\kms}{{\rm km}\,{\rm s}^{-1}}
\newcommand{\cms}{{\rm cm}^{-2}}
\newcommand{\cmc}{{\rm cm}^{-3}}

\newcommand{\Zsolar}{\;{\rm Z}_{\odot}}
\newcommand{\msolar}{{\rm M}_{\odot}}
\newcommand{\msolaryr}{{\rm M}_{\odot} {\rm yr}^{-1}}

\newcommand{\gad}{{\sc Gadget-3}}

\newcommand{\CII}{{\hbox{C\,{\sc ii}}}}
\newcommand{\CIII}{{\hbox{C\,{\sc iii}}}}
\newcommand{\CIV}{\hbox{C\,{\sc iv}}}

\newcommand{\SiI}{{\hbox{Si\,{\sc i}}}}
\newcommand{\SiII}{{\hbox{Si\,{\sc ii}}}}
\newcommand{\SiIII}{{\hbox{Si\,{\sc iii}}}}
\newcommand{\SiIV}{\hbox{Si\,{\sc iv}}}
\newcommand{\SiV}{\hbox{Si\,{\sc v}}}
\newcommand{\SiVI}{\hbox{Si\,{\sc vi}}}

\newcommand{\NV}{\hbox{N\,{\sc v}}}
\newcommand{\OI}{\hbox{O\,{\sc i}}}

\newcommand{\OIII}{\hbox{O\,{\sc iii}}}

\newcommand{\OVI}{\hbox{O\,{\sc vi}}}
\newcommand{\OVII}{\hbox{O\,{\sc vii}}}

\newcommand{\HI}{{\hbox{H\,{\sc i}}}}

\newcommand{\MgI}{{\hbox{Mg\,{\sc i}}}}
\newcommand{\MgII}{{\hbox{Mg\,{\sc ii}}}}

\newcommand{\nh}{{n_{\rm H}}}

\newcommand{\apj}{{ApJ}}
\newcommand{\apjs}{{ApJS}}

\newcommand{\mnras}{{MNRAS}}

\begin{document}
\title[Low metal ions in the simulated CGM]{The multiphase circumgalactic medium traced by low metal ions in EAGLE zoom simulations}
\author[B. D. Oppenheimer et al.]{
\parbox[t]{\textwidth}{\vspace{-1cm}
Benjamin D. Oppenheimer$^{1}$\thanks{benjamin.oppenheimer@colorado.edu}, Joop~Schaye$^{2}$, Robert A. Crain$^{3}$, Jessica K. Werk$^{4}$,  Alexander J. Richings$^{5}$}\\\\  
$^1$CASA, Department of Astrophysical and Planetary Sciences, University of Colorado, 389 UCB, Boulder, CO 80309, USA\\
$^2$Leiden Observatory, Leiden University, P.O. Box 9513, 2300 RA, Leiden, The Netherlands\\
$^3$Astrophysics Research Institute, Liverpool John Moores University, 146 Brownlow Hill, Liverpool, L3 5RF, UK\\
$^4$University of Washington, Department of Astronomy, Seattle, WA, USA\\
$^5$Department of Physics and Astronomy and CIERA, Northwestern University, 2145 Sheridan Road, Evanston, IL 60208, USA\\
}
\maketitle

\pubyear{2017}

\maketitle

\label{firstpage}

\begin{abstract}

  We explore the circumgalactic metal content traced by commonly
  observed low ion absorbers, including $\CII$, $\SiII$, $\SiIII$,
  $\SiIV$, and $\MgII$.  We use a set of cosmological hydrodynamical
  zoom simulations run with the EAGLE model and including a
  non-equilibrium ionization and cooling module that follows 136 ions.
  The simulations of $z\approx 0.2$ $L^*$
  ($M_{200}=10^{11.7}-10^{12.3} \msolar$) haloes hosting star-forming
  galaxies and group-sized ($M_{200}=10^{12.7}-10^{13.3} \msolar$)
  haloes hosting mainly passive galaxies reproduce key trends observed
  by the COS-Halos survey-- low ion column densities show 1) little
  dependence on galaxy specific star formation rate, 2) a patchy
  covering fraction indicative of $10^4$ K clouds with a small volume
  filling factor, and 3) a declining covering fraction as impact
  parameter increases from $20-160$ kpc.  Simulated $\SiII$, $\SiIII$,
  $\SiIV$, $\CII$, and $\CIII$ column densities show good agreement
  with observations, while $\MgII$ is under-predicted.  Low ions trace
  a significant metal reservoir, $\approx 10^8 \msolar$, residing
  primarily at $10-100$ kpc from star-forming and passive central
  galaxies.  These clouds tend to flow inwards and most will accrete
  onto the central galaxy within the next several Gyr, while a small
  fraction are entrained in strong outflows.  A two-phase structure
  describes the inner CGM ($<0.5 R_{200}$) with low-ion metal clouds
  surrounded by a hot, ambient medium.  This cool phase is separate
  from the $\OVI$ observed by COS-Halos, which arises from the outer
  CGM ($>0.5 R_{200}$) tracing virial temperature gas around $L^*$
  galaxies.  Physical parameters derived from standard
  photo-ionization modelling of observed column densities
  (e.g. aligned $\SiII/\SiIII$ absorbers) are validated against our
  simulations.  Our simulations therefore support previous ionization
  models indicating that cloud covering factors decline while
  densities and pressures show little variation with increasing impact
  parameter.
\end{abstract}


\section{Introduction}  

The circumgalactic medium (CGM) is thought to contain significant
reservoirs of baryons and metals outside of galaxies, extending to the
virial radius and beyond \citep[e.g.][]{che10,tum11,sto13}.
Absorption line spectroscopic observations by the Cosmic Origins
Spectrograph (COS) on the {\it Hubble Space Telescope} allow the study
of the CGM around galaxies at redshift $z\la 0.5$, where it is
easier to characterize the galaxies' properties observationally,
including their stellar mass, star formation rates (SFRs), and
morphologies.  The far-ultraviolet (FUV) spectral range of the COS
instrument (1100-1700\AA) covers numerous electronic transition lines
of metal species, including $\CII$, $\CIII$, $\CIV$, $\SiII$,
$\SiIII$, $\SiIV$, and $\OVI$ that can probe the physical state of the
gas around galaxies in the evolved Universe.

The COS-Halos survey \citep{tum13} exploited the full FUV spectral
range of COS, targeting a series of $44$ $z\approx 0.2$ galaxies
spanning stellar masses $M_*=10^{9.6}-10^{11.3} \msolar$ to explore
the CGM properties out to an impact parameter, $b=160$ kpc.  The
galaxies were selected to be either ``blue'' or ``red,'' where the
blue sample comprises star-forming galaxies and the red sample
comprises passive galaxies with little detectable star formation.
Throughout we define the COS-Halos blue and red samples as galaxies
with specific star formation rates (sSFR$\equiv$SFR$/M_*$) greater
than or less than $10^{-11}$ yr$^{-1}$, respectively.  \citet{tum11}
showed that this division in galaxy properties is reflected in the CGM
properties probed by $\OVI$ with the blue star-forming sample showing
significantly higher $\OVI$ column densities ($N_{\OVI}$) than the red
passive sample.

However, the low metal ions do not show the same behaviour as
$\OVI$. Firstly, unlike $\OVI$, the low ions do not show an obvious
dependence on sSFR \citep[][hereafter W13]{wer13}.  Secondly, W13
observed a large scatter in the column densities of ions such as
$\CII$, $\CIII$, $\SiII$, $\SiIII$, and $\MgII$.  W13 argued that the
large dispersion in low ion absorption strengths suggests that the
cool CGM is patchy in nature and hence spans a large range of
densities and/or ionization conditions.  This contrasts with $\OVI$,
which shows a significantly smaller spread in column densities around
the blue star-forming galaxies \citep{pee14}.  A third difference is
that the covering fractions of low ions decline at larger $b$ when
splitting the blue galaxy sample into two impact parameters bins
divided at $b=75$ kpc (W13).  $\OVI$ shows a much smaller decline in
column density and covering fraction with impact parameter
\citep{tum11}.  Throughout this paper, we use the general term ``low''
ion for any metal ion that is not $\OVI$, even though $\CIV$ and
$\SiIV$ are usually considered intermediate ions (e.g. W13).

\citet[][hereafter W14]{wer14} performed photo-ionization modelling
using CLOUDY \citep{fer98} to derive the physical properties traced by
$\HI$ and low metal ions believed to trace temperatures $T\sim 10^4$
K.  W14 found the gas density as a function of impact parameter to
decline from a hydrogen number density $\nh \sim 10^{-3} \cmc$ inside
$b \sim 30$ kpc to $\sim 10^{-4} \cmc$ at $b \ga 100$ kpc.  A
two-phase model based on \citet{mal04} with cool $T\sim 10^4$ K clouds
embedded in a hot $T\sim 10^6$ K halo medium is in tension with these
derived physical parameters.  This model assumes hydrostatic
equilibrium with a \citet[][NFW]{nav97} dark matter halo potential
within a cooling radius, and predicts cool CGM densities more than
$100\times$ higher than inferred from the COS-Halos observations when
combined with single-phase CLOUDY models.  However, the \citet{mal04}
model does not account for mechanical or thermal superwind feedback
imparted by star formation (SF)-driven or Active Galactic Nuclei (AGN)
feedback.

Cosmological hydrodynamic simulations that reproduce the observed
properties of galaxies require superwind feedback to eject baryons and
metals from galaxies into the CGM and intergalactic medium (IGM) to
reduce the efficiency of galactic stellar build-up
\citep[e.g.][]{spr03b, opp10, sch10}.  A consequence of metal-enriched
material leaving galaxies is an enriched IGM/CGM
\citep[e.g.][]{agu01a, the02, cen06b, opp06, wie10, smi11}.  The
Evolution and Assembly of GaLaxies and their Environments (EAGLE)
simulation project calibrated the sub-resolution prescriptions for SF
and AGN feedback to reproduce the observed $z\approx 0.1$ galactic
stellar mass function, as well as the galactic disk size and
super-massive black hole-$M_*$ relations \citep[][hereafter S15; Crain
  et al. 2015]{sch15}. Because observations of the CGM and IGM were
not used to calibrate the EAGLE simulations, the properties of gas
outside galaxies are genuine predictions of the model.

\citet[][hereafter O16]{opp16} integrated the non-equilibrium (NEQ)
ionization and dynamical cooling module introduced in \citet{opp13a}
into the EAGLE simulation code to trace the evolution of 136 ions of
11 elements.  O16 ran a set of 20 zoom simulations of individual
galactic haloes, 10 of which host blue, star-forming galaxies and have
virial masses $\sim 10^{12} \msolar$ and another 10 haloes at $\sim
10^{13} \msolar$ most of which host red, passive galaxies.  They
turned on the NEQ module at low redshift to follow non-equilibrium
effects over the redshift range of COS-Halos galaxies.  They found
that for a COS-Halos-like sample, $\OVI$ is strongest around blue
galaxies, because the temperatures of the virialized gas in their host
haloes overlap with the $3\times 10^5$ K collisional ionization
temperature of $\OVI$.  $\OVI$ is less strong around red COS-Halos
galaxies, because their host halo virial temperatures exceed $10^6$ K
resulting in CGM oxygen being promoted to $\OVII$ and above.  O16
argued that the correlation between circumgalactic $N_{\OVI}$ and
galactic sSFR observed by \citet{tum11} is not causal, but reflects
the increasing ionization state of oxygen with virial mass and
temperature.

The same COS-Halos sight lines that show $\OVI$ often also show low
metal ions and $\HI$ \citep{tho12}, implying that the CGM is
multiphase.  The EAGLE NEQ zoom hydrodynamic simulations are
well-suited for a study of the multiphase CGM.  Our zooms
self-consistently follow the nucleosynthetic production of heavy
elements in stars, their propagation out of galaxies due to superwind
feedback, and the detailed non-equilibrium atomic processes setting
the ionization states in the CGM.  Here we extend the work of O16 to
the COS-Halos low metal ions in the same zooms that were used by O16
to explain the $N_{\OVI}$-sSFR correlation.  We mention that the
recent work of \citet{opp17} includes fluctuating AGN radiation added
to one of our zooms, which results in enhanced $\OVI$ column densities
around COS-Halos galaxies.  We discuss this work throughout our
investigation of low ions, but note that these ions, unlike $\OVI$,
are not nearly as strongly affected by fluctuating AGN radiation.
 
The paper is organized as follows.  We describe our simulations and
review our non-equilibrium module in \S\ref{sec:sims}.  We describe how
the low ion-traced CGM changes as a function of halo mass in
\S\ref{sec:halotrends}, and then compare directly to COS-Halos
observations in \S\ref{sec:obs}.  The physical and evolutionary state
of low ions is addressed in \S\ref{sec:phys} with discussions of metal
masses, physical gas parameters, evolution of low ion-traced gas from
$z=0.2\rightarrow 0.0$, and ion ratios.  We summarize in
\S\ref{sec:summary}.  Resolution and non-equilibrium effects are
explored in the Appendix, as well as statistical methods.

\section{Simulations} \label{sec:sims}  

We briefly describe the simulations in this section, and refer the
reader to \S2 of O16 for further details.  We employ the EAGLE
hydrodynamic simulation code described in S15, which is an extensively
modified version of the N-body+Smoothed Particle Hydrodynamic (SPH)
\gad~code last described in \citet{spr05}.  We assume the
\citet{pla14} cosmological parameters adopted in EAGLE simulations:
$\Omega_{\rm m}=0.307$, $\Omega_{\Lambda}=0.693$, $\Omega_b=0.04825$,
$H_0= 67.77$ $\kmsmpc$, $\sigma_8=0.8288$, and $n_{\rm s}=0.9611$.
EAGLE uses the \citet{hopk13} pressure-entropy SPH formulation
applying a C2 \citet{wen95} 58-neighbour kernel along with several
other hydrodynamic modifications collectively referred to as
``Anarchy'' \citep[Appendix A of S15 and ][]{schal15}.  The EAGLE code
includes subgrid prescriptions for radiative cooling \citep{wie09a},
star formation \citep{sch08}, stellar evolution and chemical
enrichment \citep{wie09b}, and superwind feedback associated with star
formation \citep{dal12} and black hole growth \citep[S15;][]{ros15}.

EAGLE provides an ideal testbed for the study of the CGM, because it
successfully reproduces an array of galaxy observables
\citep[e.g. S15;][]{fur15, fur17, tra15, bah16, seg16a} in a model
that explicitly follows the hydrodynamics.  Even though the EAGLE
model was not calibrated on observations of the IGM/CGM, EAGLE
simulations show broad but imperfect agreement with absorption line
statistics probing $\HI$ \citep{rah15} and metal ions \citep{rah16,
  tur16, tur17}, and the $\OVI$ bimodality observed around COS-Halos
galaxies (O16).

\subsection{Non-equilibrium network} \label{sec:NEQcode}

The NEQ module \citep{ric14a}, integrated into the EAGLE
\gad~simulation code by O16, explicitly follows the reaction network
of 136 ionization states for the 11 elements that significantly
contribute to the cooling (H, He, C, N, O, Ne, Si, Mg, S, Ca, \& Fe)
plus the electron density of the plasma.  Our reaction network is
described in \citet{opp13a}.  It includes radiative and di-electric
recombination, collisional ionization, photo-ionization, Auger
ionization, and charge transfer.  Cooling is summed ion-by-ion
\citep{gna12,opp13a} over all 136 ions.  The method has been verified
to reproduce results obtained from other codes and is interchangeable
with the equilibrium elemental cooling tables of \citet{wie09a} that
were used in other EAGLE runs.

Our EAGLE zooms assume an interstellar medium (ISM), defined as gas
having non-zero SFR, with a single phase where we do not follow the
NEQ behaviour and instead use equilibrium lookup tables tabulated as
functions of density assuming $T=10^4$ K.  This makes little
difference for most ions, but it can affect the balance between the
lowest ion states such as $\SiI$ and $\SiII$ or $\MgI$ and $\MgII$.
However, we concern ourselves with non-star-forming CGM gas throughout
unless specifically noted otherwise.  Metal enrichment from stars onto
gas particles releases new metals in their ground-state ions.
However, the vast majority of the enrichment occurs in ISM gas where
ISM equilibrium tables are used.

We run simulations using the standard ``equilibrium'' EAGLE code to
low redshift and then turn on the NEQ network at $z\leq 0.5$ as
described in \S2.3 of O16.  The only difference with standard EAGLE
runs (S15), which use kernel-smoothed metal abundances, is that we use
particle-based metal abundances in all EAGLE equilibrium runs and
particle-based ion abundances in NEQ runs.  Appendix B1 of O16 found
that circumgalactic $\OVI$ is nearly unchanged when using
particle-based instead of kernel-smoothed metallicities, but stellar
masses decline by 0.1 dex when using particle-based metallicities.

\subsection{Runs} \label{sec:runs} 

We use the set of zoom simulations listed in Table 1 of O16, but we
also add a $12.5$ Mpc simulation periodic volume described in detail
below.  Our main resolution is the {\it M5.3} resolution of O16
corresponding to an initial SPH particle mass $m_{\rm SPH} = 2.2\times
10^5 \msolar$, using the notation {\it M}[log($m_{\rm SPH}/\msolar$)].
This resolution has a Plummer-equivalent softening length of 350
proper pc at $z<2.8$, and 1.33 comoving kpc at $z>2.8$.

\noindent{\bf Zooms: } Twenty zoom simulations centered on haloes with
mass $M_{200} = 10^{11.8}- 10^{13.2} \msolar$, where $M_{200}$ is the
mass within a sphere within which the mean internal density is
$200\times$ the critical overdensity.  Ten haloes corresponding to
``$L^*$'' masses ($M_{200}=10^{11.7}-10^{12.3} \msolar$) were selected
from the EAGLE Recal-L025N0752 simulation and ten haloes corresponding
to ``group'' masses ($M_{200}=10^{12.7}-10^{13.3} \msolar$) were
selected from the Ref-L100N1504 simulation.  Additionally, several
zooms contain ``bonus'' haloes that were verified to reside completely
within the region resolved with the high-resolution SPH and dark
matter particles.

We only activate the NEQ module at low redshift in order to reduce
computational cost, and because the NEQ effects on CGM ionization
levels are short-lived compared to the Hubble timescale.  $L^*$ (group)
zooms are run using the NEQ module beginning at $z=0.503$ ($0.282$).
We use outputs of zooms at $z=0.250$, $0.205$, and $0.149$ in our
analysis here.  O16 used additional outputs at $z=0.099-0.0$ to obtain
a wider range of galaxy properties to simulate COS-Halos, but we
decided not to use these additional outputs since they do not overlap
with COS-Halos redshifts and they do not statistically alter the
simulation results.  Standard equilibrium EAGLE runs are also run to
$z=0$ and we use these runs at $z=0.20$ for comparison to NEQ runs in
\S\ref{sec:mods}.

\noindent{\bf Periodic Volume: } We add a 12.5 Mpc, $376^3$ SPH + DM
particle simulation to our analysis here, which is a Recal-L012N0376
simulation using EAGLE terminology.  This simulation was run in NEQ
from $z=0.503\rightarrow 0.0$, and we use outputs at $z=0.351$,
$0.250$, $0.205$, and $0.149$.  This volume contains several
$L^*$/group halos, which we add to our halo sample, plus a large range
of haloes with $M_{200}<10^{11.7} \msolar$, which we term ``sub-$L^*$''
haloes.  This allows us to simulate the three lower mass COS-Halos
galaxies at $M_*<10^{9.7} \msolar$, which were removed from the
comparison in O16.

\subsection{Isolation criteria}\label{sec:iso}

For each central galaxy, we test whether it is defined as ``isolated''
using similar criteria as those used to select the COS-Halos sample.
However, there exists some ambiguity in how isolated the COS-Halos
galaxies truly are.  \citet{tum13} reports that COS-Halos galaxies are
``the most luminous galaxy within 300 kpc of the QSO sightline at its
redshift.''  The spectroscopic and photometric galaxy field follow-up
of \citet{wer12} found many $L>0.1 L^*$ galaxies often within 160 kpc
of the targeted COS-Halos galaxy.  The initial COS-Halos galaxy
selection used only photometric redshifts to select ``isolated''
galaxies, so it is not surprising that deeper follow-up has resulted
in the discovery of neighbouring galaxies at similar redshifts as
described in detail in W13.

We therefore make two isolation criteria: 1) the ``stringent''
criteria that there should not be any galaxies within $b=300$ kpc
having $M_*>2\times 10^{10} \msolar$ applied in O16, and 2) the
``loose'' isolation criteria that reduces $b$ to $100$ kpc but also
reduces the minimum stellar mass to $M_*>10^{10} \msolar$.  We project
all central galaxies in three directions ($x$, $y$, \& $z$) and test
the criteria in each direction.  Nearly all $L^*$ galaxies satisfy
both isolation criteria, while over half of group galaxy directions
are thrown out using the stringent criteria, which reduces to $\approx
20\%$ using the loose criteria.  O16 used the stringent criteria for
$\OVI$, but we favor the loose criteria for this work given the
follow-up of \citet{wer12}.  This is because the stringent criteria
result in the elimination of about half of the prospective passive
COS-Halos targets from the COS-Halos sample selection, and such a cut
was not applied to that survey (J. Tumlinson, private communication).

We will show in \S\ref{sec:obs} that the chosen isolation criteria
make a more significant difference for low ions than for $\OVI$ around
group galaxies.  $N_{\OVI}$ increased by $\approx 0.2$ dex upon
eliminating the stringent isolation criteria with no isolation
criteria around group galaxies (O16).  The fit to the COS-Halos $\OVI$
using the new loose criteria is essentially identical to the stringent
criteria, because the passive galaxy sight lines are mostly upper
limits.  In general, we use the loose isolation criteria in our mock
observational samples, however we will compare to the stringent
isolation criteria in certain instances.

\section{Low metal ions in the circumgalactic medium} \label{sec:halotrends}

We begin our presentation of results by considering the CGM as traced
by different ions within 300 kpc of galaxies as a function of halo
mass.  Our purpose is to provide an understanding of how observations
of column density as a function of impact parameter depend on host
halo mass before we compare directly to COS-Halos data in
\S\ref{sec:obs} and consider the physical conditions of the cool CGM
traced by low ions in \S\ref{sec:phys}.  We mainly concentrate on low
silicon ions.  W13 showed that the three main differences between low
ions and $\OVI$, which we studied in O16, are that the former have 1)
little dependence on sSFR, 2) a larger range in column density
indicating a patchier covering fraction, and 3) a more strongly
declining covering fraction at large impact parameters for blue
galaxies.

\begin{figure*}
\includegraphics[width=0.32\textwidth]{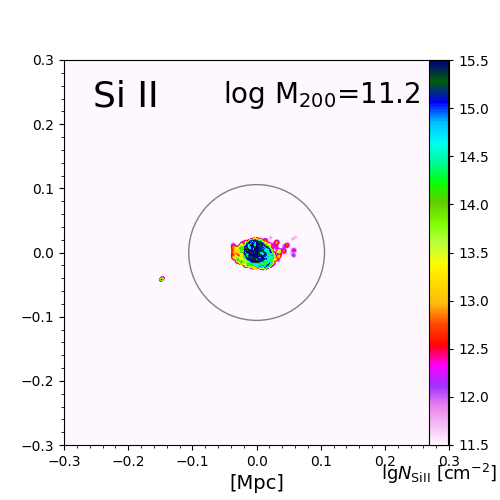}
\includegraphics[width=0.32\textwidth]{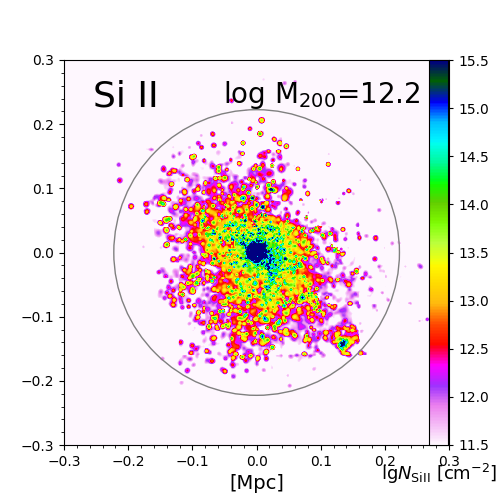}
\includegraphics[width=0.32\textwidth]{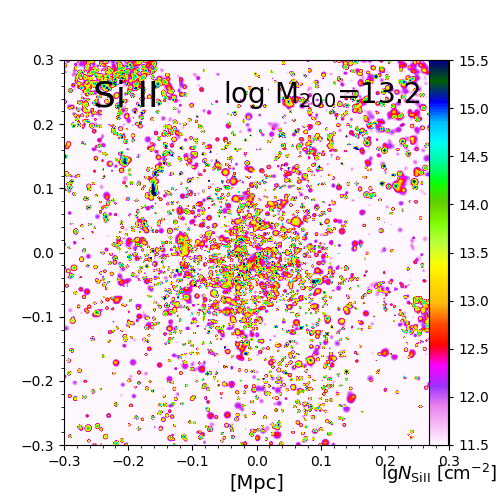}
\includegraphics[width=0.32\textwidth]{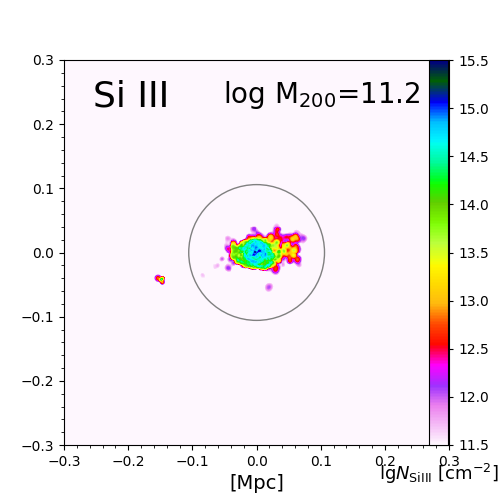}
\includegraphics[width=0.32\textwidth]{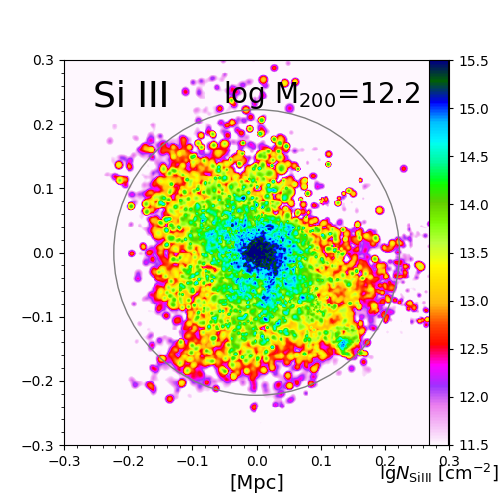}
\includegraphics[width=0.32\textwidth]{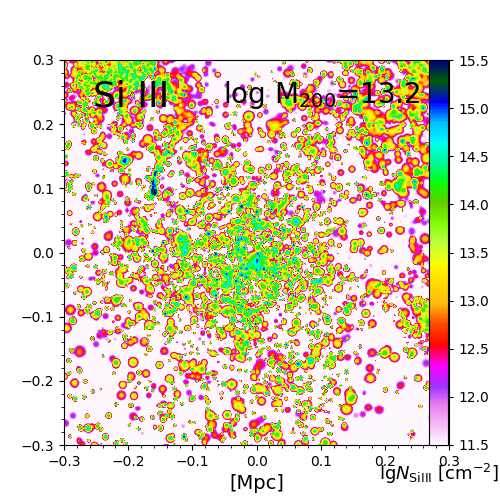}
\includegraphics[width=0.32\textwidth]{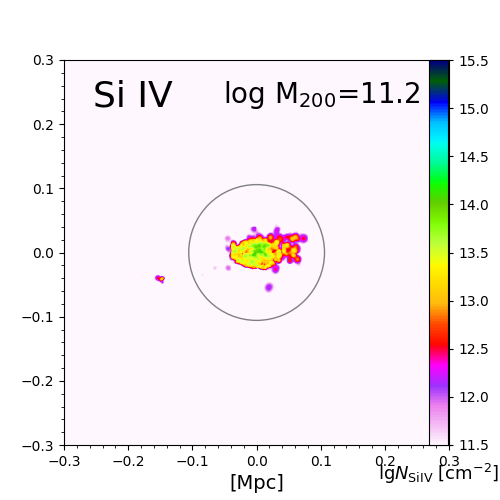}
\includegraphics[width=0.32\textwidth]{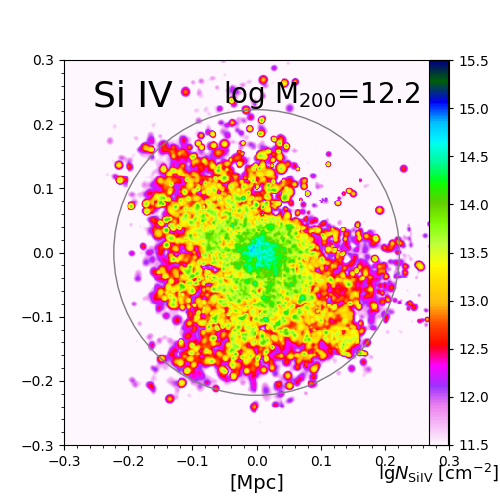}
\includegraphics[width=0.32\textwidth]{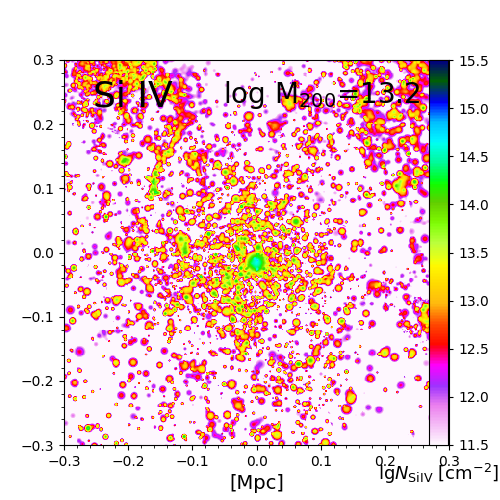}
\includegraphics[width=0.32\textwidth]{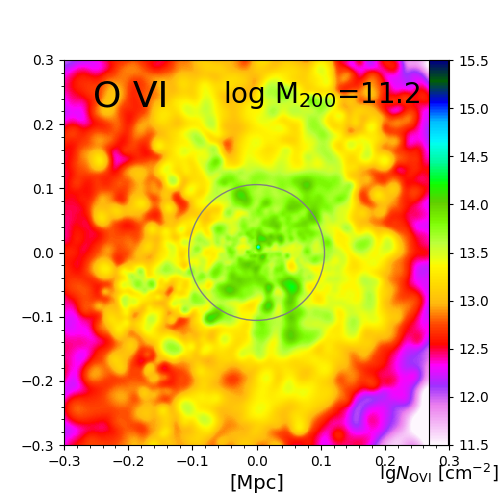}
\includegraphics[width=0.32\textwidth]{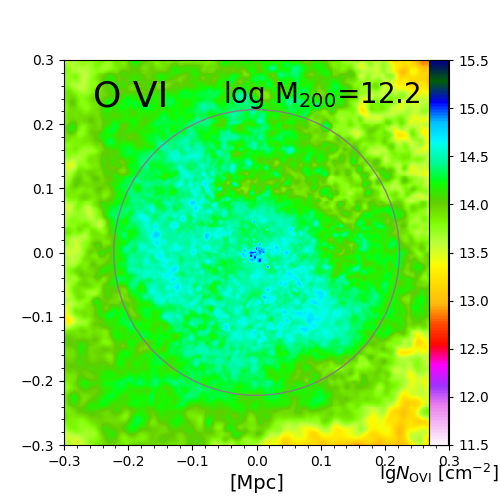}
\includegraphics[width=0.32\textwidth]{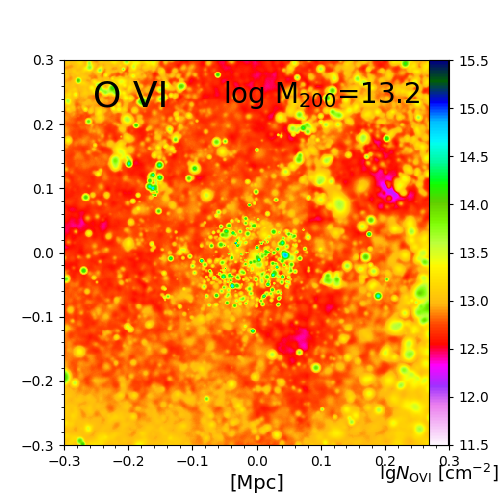}
\caption[]{Column density maps for $z=0.205$ snapshots of three haloes
  with mass $10^{11.2}$, $10^{12.2}$, and $10^{13.2} \msolar$
  representative of sub-$L^*$, $L^*$, and group-sized haloes,
  respectively, from left to right.  From top to bottom, the rows show
  $\SiII$, $\SiIII$, $\SiIV$, and $\OVI$ column densities on a
  $600\times600$ kpc grid.  Grey circles indicate $R_{200}$, which is
  too large ($486$ kpc) to appear in the group halo frame.}
\label{fig:halotrends}
\end{figure*}

Figure \ref{fig:halotrends} shows column density maps of three
$z=0.20$ haloes with masses $10^{11.2}$, $10^{12.2}$, and $10^{13.2}
\msolar$ corresponding to sub-$L^*$, $L^*$, and group galaxies
respectively.  Focusing first on the $L^*$ halo in the center column,
which we and O16 argue corresponds to the blue COS-Halos sample, we
see that the silicon species (top three rows) have patchier
distributions and are much more concentrated around the galaxy
compared to the $\OVI$ shown in the lower panel.

Moving from lower to higher halo mass for the silicon species, we see
dramatic changes. The sub-$L^*$ CGM shows silicon absorption in an
extended disky structure out to at most $b\sim 50$ kpc, while the
$L^*$ halo is covered with low ion silicon absorption out to 150-200
kpc.  The group halo shows much less low ion silicon absorption in the
central regions, but the patchy distribution extends beyond the radius
where it falls off for the $L^*$ galaxy.  The $\OVI$ by contrast is
stronger everywhere within the central 300 kpc for the $L^*$ galaxy
than for the sub-$L^*$ and group halo, owing to the O16 explanation of
the $3\times10^5$ K collisionally ionized band overlapping with the
virial temperature of a $\approx 10^{12} \msolar$ halo.

\begin{figure*}
\includegraphics[width=0.49\textwidth]{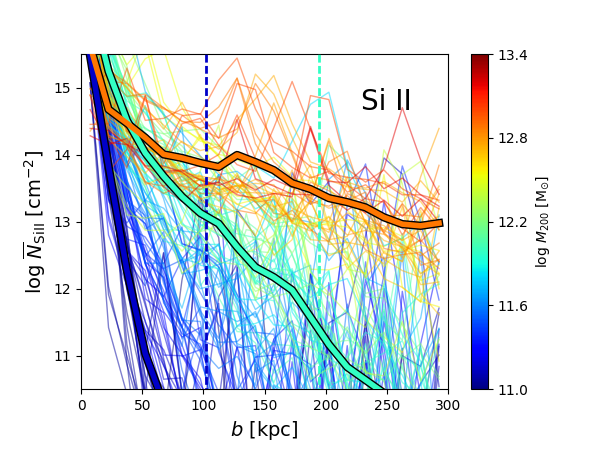}
\includegraphics[width=0.49\textwidth]{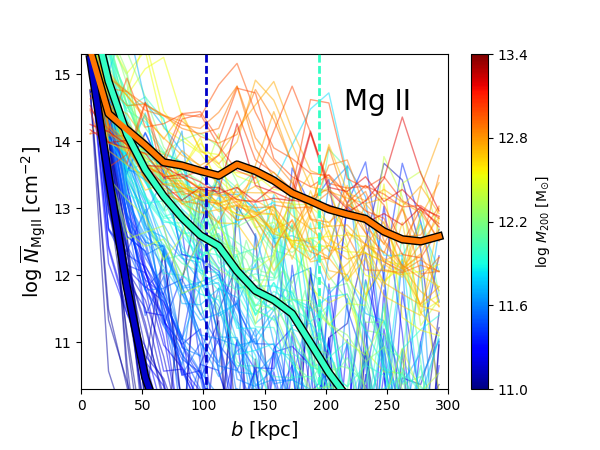}
\includegraphics[width=0.49\textwidth]{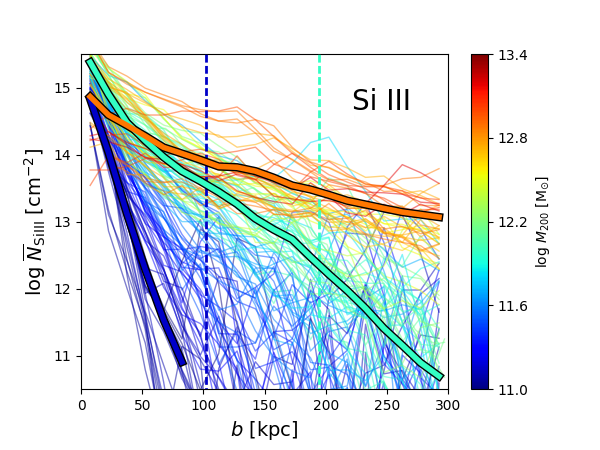}
\includegraphics[width=0.49\textwidth]{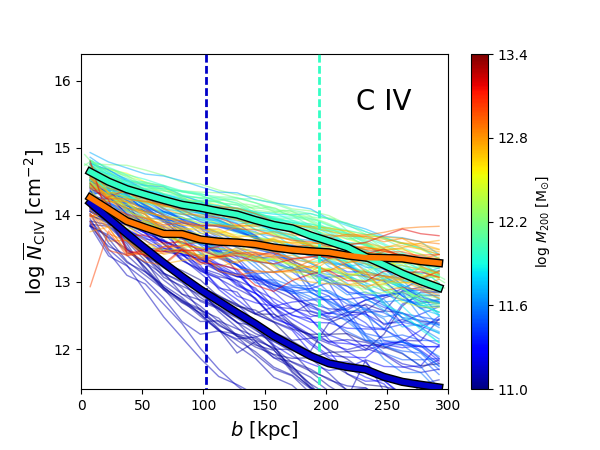}
\includegraphics[width=0.49\textwidth]{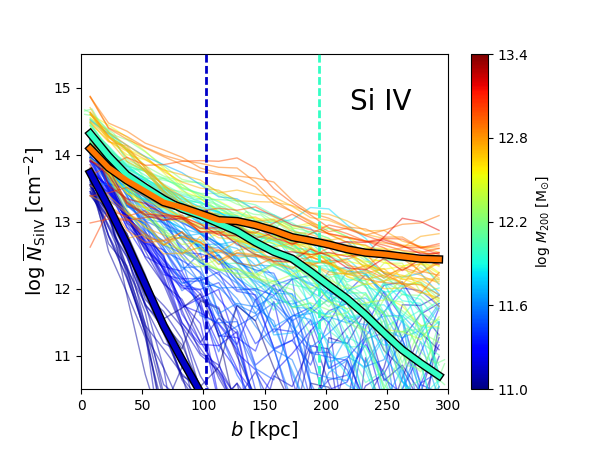}
\includegraphics[width=0.49\textwidth]{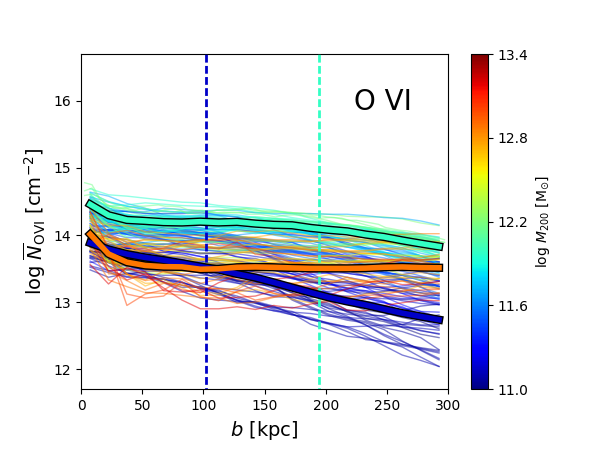}
\caption[]{Linearly averaged column densities as a function of impact
  parameter at $z=0.2$, coloured by halo mass for silicon species on
  the left ($\SiII$, $\SiIII$, \& $\SiIV$ from top to bottom) and for
  $\MgII$, $\CIV$, and $\OVI$ on the right.  Individual galaxies in
  isolated projections are shown as thin lines, and averages for
  sub-$L^*$, $L^*$, and group-sized haloes (blue
  $M_{200}=10^{11.0}-10^{11.3}\msolar$, aquamarine
  $10^{11.7}-10^{12.3}\msolar$, \& orange
  $10^{12.7}-10^{13.3}\msolar$) are shown as bordered, thick lines.
  Dashed vertical lines indicate average $R_{200}$ for the 3 samples
  (the group haloes have $R_{200}=384$ kpc). The column density range
  for each panel is scaled according to the relative abundance of each
  element in the simulation.  Hence, the relative locations of the
  curves within their panels reflect the differences in ion
  fractions.}
\label{fig:bave_Nion}
\end{figure*}

In Fig. \ref{fig:halotrends}, the silicon species look stronger in the
$L^*$ CGM than in the group CGM, but we present a more quantitative
approach in Fig. \ref{fig:bave_Nion} by plotting the linearly averaged
column densities, $\overline{N}_{\rm ion}$, as a function of impact
parameter, $b$, by taking the average $N_{\rm ion}$ in annuli of 15
kpc width.  We plot the silicon species ($\SiII$, $\SiIII$, \&
$\SiIV$) along the left column for all central galaxies that appear as
isolated in the $x$, $y$, and $z$ directions as thin lines.  We plot
averages as thick lines with black borders for sub-$L^*$
($M_{200}=10^{11.0}-10^{11.3} \msolar$, dark blue), $L^*$
($M_{200}=10^{11.7}-10^{12.3} \msolar$, aquamarine), and group
($M_{200}=10^{12.7}-10^{13.3} \msolar$, orange) subsamples.  The Si
species in the sub-$L^*$ CGM fall off rapidly, while the $L^*$ CGM has
stronger Si species inside $\approx 50$ kpc than groups, but groups
have more extended, shallower distributions of low Si ions.

We also show $\MgII$ in Fig. \ref{fig:bave_Nion} (top right panel),
which shows very similar trends as $\SiII$, although $\MgII$ has a
slightly lower ionization potential than $\SiII$ meaning that it acts
like a slightly lower ion.  We also show $\CIV$ (right middle panel),
which has a higher ionization potential than $\SiIV$, and the high ion
$\OVI$ (lower right panel).  For each species, we show a 5 dex range
in column density with the y-axis scaled to the same relative
abundance based on the atomic number density within the simulation
(e.g. the $\overline{N}_{\OVI}$ range is $1.2$ dex higher than for
$\overline{N}_{\SiII}$, $\overline{N}_{\SiIII}$, and
$\overline{N}_{\SiIV}$, because there are $\approx 10^{1.2}$ more
oxygen than silicon atoms).  This helps visualize the effect of the
ion fractions on the strengths of the various ion species.  For
example, the average group has more Si in $\SiII$ than O in $\OVI$,
since the bordered orange line is higher for $\overline{N}_{\SiII}$
than for $\overline{N}_{\OVI}$ at all radii.  It may be surprising
that $\SiII$, which traces $\approx 10^4$ K gas, is relatively more
abundant than $\OVI$, which primarily traces $>10^5$ K gas, in a group
whose CGM is dominated by $\ga 10^6$ K gas.

The progression of ions from lowest ($\MgII$) to highest ($\OVI$)
shows the following trends: 1) lower ions have less extended
distributions, 2) lower ions have significantly more scatter, even
when plotting quantities binned in 15 kpc-wide annuli, and 3) the
lower the ion, the smaller the impact parameter within which the $L^*$
column densities exceed the group column densities.  Several of the
trends visible in Fig. \ref{fig:bave_Nion} have been observed.  $\CIV$
declines faster around sub-$L^*$ galaxies in COS-Dwarfs \citep{bor14}
than around more massive galaxies ($M_*>10^{9.5} \msolar$) as observed
by \citet{bur16} (cf. blue and aquamarine bordered lines in right
middle panel).  \citet{bur16} also see a decline in $\CIV$ detection
for higher halo masses ($M_{200}>10^{12.5} \msolar$), especially
inside 160 kpc (cf. orange \& aquamarine bordered lines).
\citet{lia14} and \citet{bort16} observed virtually no $\SiIII$ beyond
$\approx 0.7-0.8\times$ the virial radius around their samples dominated
by $L^*$ halo mass objects, which agrees with the steep decline seen
in $\SiIII$ in our $L^*$ sample (left middle panel).  Local ionizing
radiation from galaxies, not included in these simulations, could
reduce low ion column densities preferentially in the inner CGM as we
discuss in \S\ref{sec:mods}.

Finally, the $\OVI$ averages in the lower right panel of
Fig. \ref{fig:bave_Nion} show remarkably similar impact parameter
profiles inside 150 kpc for sub-$L^*$ and group galaxies, but the
origins of $\OVI$ are very different.  As discussed in O16, sub-$L^*$
galaxies have photo-ionized $\OVI$ in their $<10^5$ K CGM, while group
galaxies have very low $\OVI$ fractions in their collisionally ionized
$>10^6$ K CGM.  However, individual $\OVI$ sight line measurements are
predicted to be quite different with sub-$L^*$ galaxies showing less
scatter in the $\OVI$ column densities, and group galaxies showing
more scatter with significantly lower median $\OVI$ column densities
(cf. lower left and right panels of Fig. \ref{fig:halotrends}).

\subsection{The effect of neighbouring galaxies} \label{sec:neighbours}

It may be counter-intuitive that low ions are more abundant around
hotter gas haloes, which mostly host passive galaxies with little
star-formation.  For $\CII$, $\CIII$, $\MgII$, $\SiII$, $\SiIII$, and
$\SiIV$, group column densities exceed $L^*$ column densities at every
impact parameter $>90$ kpc (cf. orange and aquamarine bordered lines
in Fig. \ref{fig:bave_Nion} panels).  We reconsider the stringent
isolation criteria used in O16 to check how the loose isolation
criteria we use in this figure differs.  The stringent criteria
results in a decline of $\approx 0.2-0.3$ dex between $20-160$ kpc for
low ions, meaning that neighbouring galaxies at $b=100-300$ kpc can
increase low ion column densities by a factor of $\approx 1.5-2$.

\begin{figure*}
\includegraphics[width=0.49\textwidth]{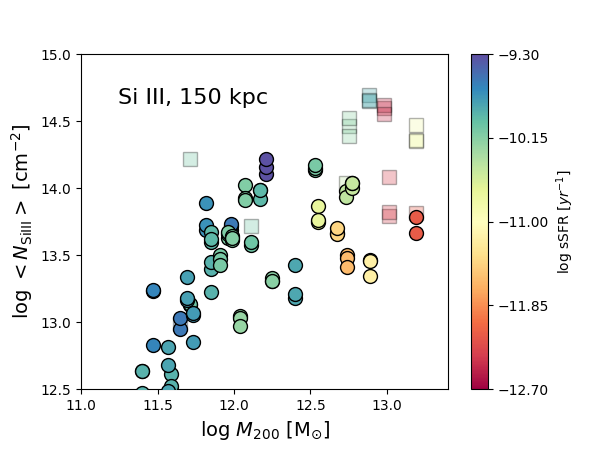}
\includegraphics[width=0.49\textwidth]{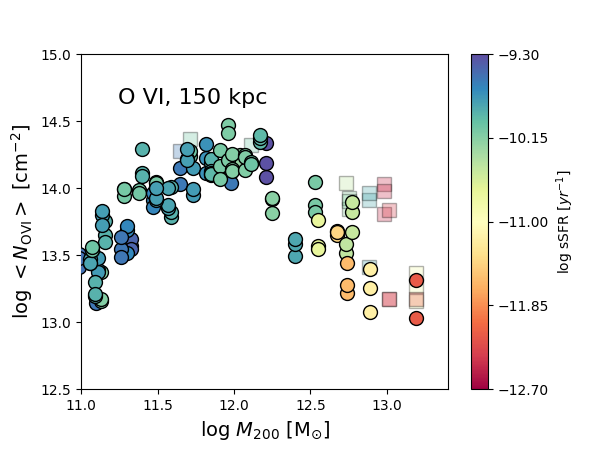}
\caption[]{Aperture column densities of $\SiIII$ (left) and $\OVI$
  (right) averaged within 150 kpc of the central $z=0.2$ galaxy and
  plotted as a function of halo mass.  Solid circles indicate
  stringently isolated galaxies and transparent squares indicate
  non-isolated galaxies.  Colour indicates sSFR.}
\label{fig:mhalo_Nave}
\end{figure*}

Figure \ref{fig:mhalo_Nave} shows the 150 kpc ``aperture'' column
densities for $\SiIII$ and $\OVI$ in the $x$, $y$, and $z$ directions
(there are three data points for each halo), where the aperture column
density is defined as

\begin{equation} \label{equ:apcol} 
\langle N \rangle_{b} =  \frac{\sum\limits_{<b} N(x,y) dx^2}{\pi b^2} \cms
\end{equation}

\noindent where $b$ is the impact parameter and $dx$ is the pixel size
($dx\ll b$).  The $\SiIII$ aperture columns, $\langle
N_{\SiIII}\rangle_{150}$, increase faster with $M_{200}$ from
sub-$L^*$ to $L^*$ haloes than for $\OVI$.  However, while $\langle
N_{\OVI}\rangle_{150}$ declines from $L^*$ to group haloes, as
extensively detailed in O16, $\langle N_{\SiIII}\rangle_{150}$ shows
no decline and a much larger scatter.

Stringently isolated projections, shown as solid circles, have
$\langle N_{\SiIII}\rangle_{150}$ values that are similar between
groups and $L^*$ haloes, but non-isolated counterparts with
neighbouring galaxies within 300 kpc, plotted as transparent squares,
indicate a separate branch where $\langle N_{\SiIII}\rangle_{150}$
increases from $L^*$ to group haloes.  Thus, the typical column
density of low ions observed in COS-Halos at $b<150$ kpc would likely
show less dependence on the properties of the central galaxy than is
the case for $\OVI$.  However, Fig. \ref{fig:bave_Nion} shows that the
$N_{\rm ion}(b)$ relations for these low ions are most different
between the $L^*$ and group samples beyond $b=150$ kpc, which are not
plotted in Fig. \ref{fig:mhalo_Nave}.  Group haloes, even if isolated,
still have more low ions beyond 150 kpc than $L^*$ haloes.
  

Lastly, in Fig. \ref{fig:mhalo_Nave} we colour the $\langle
N_{\OVI}\rangle_{150}$ values by sSFR (Fig. \ref{fig:mhalo_Nave} right
panel) to show how $\OVI$ columns are driven by halo mass rather than
sSFR (O16), which does not appear to be the case for $\SiIII$.  In our
simulations, the $\OVI$ column density shows less dependence on
whether the central has neighbouring galaxies than is the case for
$\SiIII$.

\section{Comparison to COS-Halos data} \label{sec:obs}

We now compare our simulation results to the COS-Halos observational
survey using the python module named Simulation Mocker Of Hubble
Absorption-Line Observational Surveys (SMOHALOS) described in O16.
SMOHALOS creates mock COS-Halos surveys using observed impact
parameters for galaxies chosen to match the COS-Halos $M_*$ and sSFR.
We use the latest spectroscopic galaxy data \citep{wer12} and the most
recently published values of the absorption line observations (W13) in
our SMOHALOS realizations.

As also described in O16, SMOHALOS applies observational errors from
\citet{wer12} to simulated galaxy measurements using a random number
generator.  Gaussian dispersions of $0.2$ and $0.3$ dex are applied to
simulated log $M_*$ and log sSFR values, respectively.  The dispersed
simulated $M_*$ and sSFR values closest to the observed $M_*$ and sSFR
are then selected.  Stellar masses assume a \citet{cha03} initial mass
function (IMF), which requires us to reduce the stellar masses
reported by \citet{wer12} by 0.2 dex, because they assumed a
\citet{sal55} IMF.  The observed $b$ is matched by SMOHALOS through a
random number generator picking a pixel at the same $b$ in one of
three column density maps ($x$, $y$, \& $z$ projections) that
satisfies the isolation criteria.  Like O16, we do not require a
simulated galaxy to have the same redshift as the observed galaxy,
because we find little evolution over the considered redshift range
($z=0.15-0.35$).  One hundred SMOHALOS realizations are run to compare
to the 44 galaxies from COS-Halos for a total of 4400 measurements.

\begin{figure*}
\includegraphics[width=0.49\textwidth]{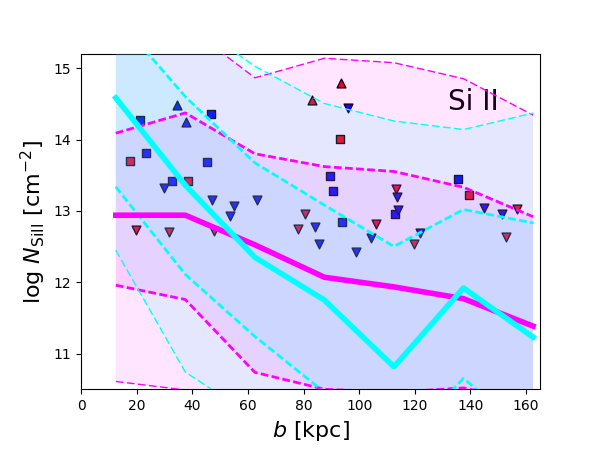}
\includegraphics[width=0.49\textwidth]{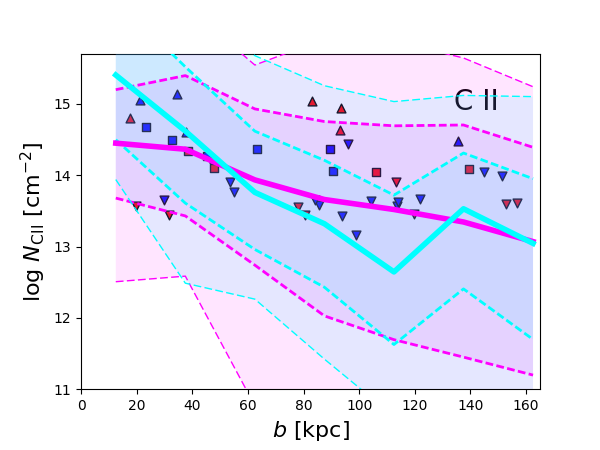}
\includegraphics[width=0.49\textwidth]{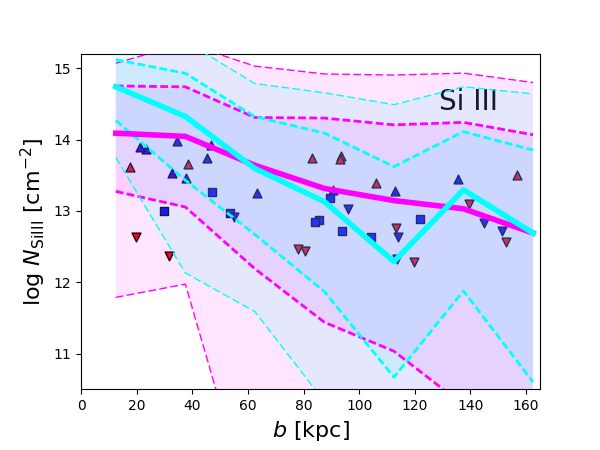}
\includegraphics[width=0.49\textwidth]{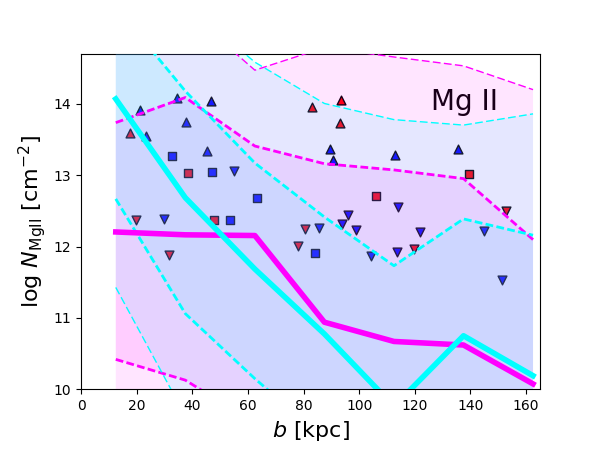}
\includegraphics[width=0.49\textwidth]{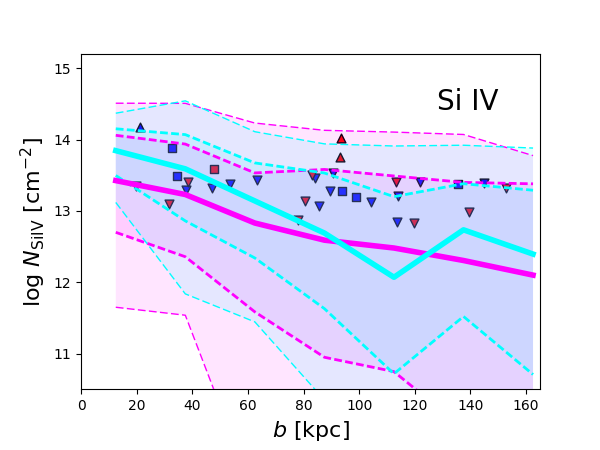}
\includegraphics[width=0.49\textwidth]{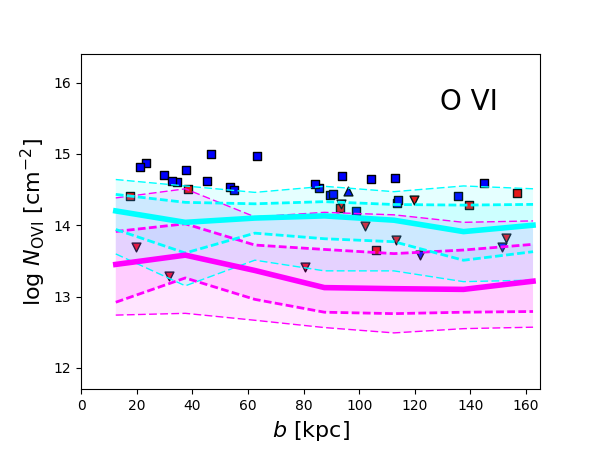}
\caption[]{Column density as a function of impact parameter for the
  COS-Halos blue and red samples (squares are detections, upside-down
  triangles are upper limits, and upwards pointing triangles are lower
  limits).  Simulated column densities from 100 SMOHALOS realizations
  are plotted as solid cyan and magenta lines for galaxies with sSFR
  higher and lower than $10^{-11}$ yr$^{-1}$.  One and 2 $\sigma$
  dispersions are indicated by thick and thin dashed lines,
  respectively.  Three silicon ions ($\SiII$, $\SiIII$, \& $\SiIV$,
  left panels) are shown along with $\CII$, $\MgII$, and $\OVI$ (right
  panels).  Compared to $\OVI$, the simulated low ions have less
  dependence on sSFR, as indicated by overlapping cyan and magenta
  regions, a patchier distribution, as indicated by larger
  dispersions, and column densities that decline faster at larger
  impact parameters.  A comparison of the simulations to the
  observations is difficult owing to the dominance of upper and lower
  limits (see Figs. \ref{fig:KM_all} and \ref{fig:KM_ssfr} instead).}
\label{fig:b_SMOHALOS_prob}
\end{figure*}

Figure \ref{fig:b_SMOHALOS_prob} shows the COS-Halos observations
(W13) for $\SiII$, $\SiIII$, $\SiIV$, $\CII$, $\MgII$, and $\OVI$ as a
function of impact parameter.  Blue and red symbols indicate sSFR
greater than and less than $10^{-11} $ yr$^{-1}$, respectively.
Squares indicate detections, upside-down triangles indicate upper
limits for non-detections, and upwards pointing triangles indicate
lower limits for saturated lines.  The median SMOHALOS column density
as a function of impact parameter for the blue and red samples, using
a division at sSFR$= 10^{-11}$ yr$^{-1}$, are shown as cyan and
magenta lines, respectively.  One and 2 $\sigma$ dispersions are
indicated by thick and thin dashed lines where the simulated absorbers
are ``perfect'' data (i.e. exact column densities, no upper or lower
limits).

As in Fig. \ref{fig:bave_Nion}, all $y$-axis ranges are scaled to the
same relative abundances.  It is difficult to assess the agreement
with the observations from this plot alone given the dominance of
lower and upper limits in the data.  There are no low ion detections
outside the $2$-$\sigma$ simulated SMOHALOS bands, unlike is the case
for $\OVI$, which is $2-3\times$ too weak in our simulations (O16).
The simulations show the same contrasts between low ions and $\OVI$ as
observed in COS-Halos: 1) less dependence on the sSFR as indicated by
the smaller differences between the medians, 2) a patchier
distribution as indicated by larger $1$ and $2$-$\sigma$ dispersions,
and 3) more strongly declining column densities at larger impact
parameters particularly around blue galaxies.  In the simulations, the
dispersion and the dependence on impact parameter decline going from
the lowest species ($\SiII$, $\CII$, $\MgII$) through ``intermediate''
species ($\SiIII$, $\SiIV$) up to $\OVI$.

\subsection{Survival analysis} \label{sec:surv}

\subsubsection{Application method}

To test the quality of the match between simulations and COS-Halos, we
need to account for the observed upper and lower limits, for which we
turn to survival analysis.  Survival analysis allows a statistical
interpretation of incomplete datasets where a portion of the data are
``censored.''  The Kaplan-Meier (K-M) method provides a general
one-variable, non-parametric survival statistic that produces a
maximum likelihood distribution using both uncensored (detections) and
censored (upper or lower limits) data.  While this method has been
applied to astronomical datasets including upper limits
\citep[e.g.][]{fei85} and discussed extensively in the context of
absorption line surveys in \citet{sim04}, we apply a two-sided
censored K-M estimator that we argue applies more proper treatment as
well as limitations compared to a one-sided censoring K-M estimator.

\begin{figure*}
\includegraphics[width=0.98\textwidth]{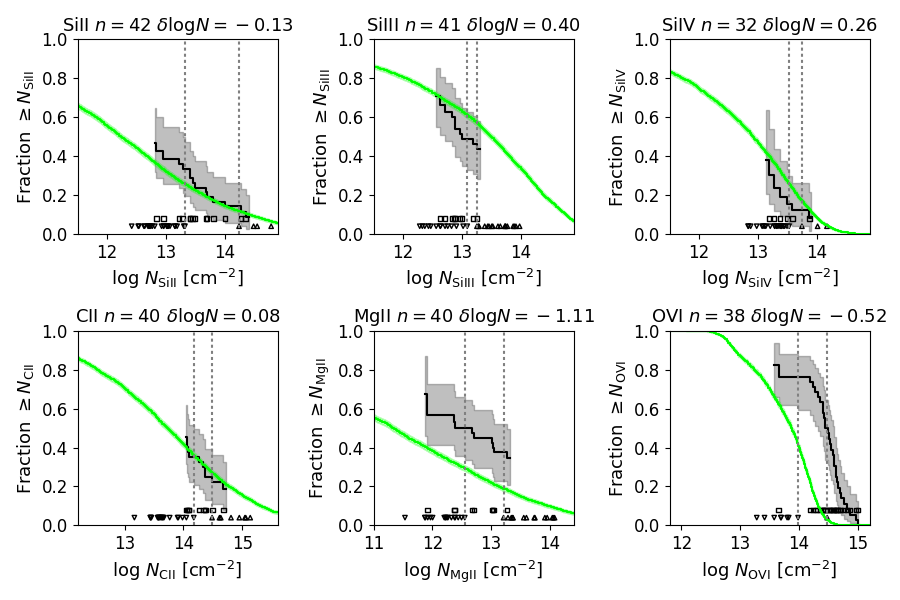}
\caption[]{Cumulative distribution functions (CDFs) of COS-Halos
  column densities for various ions (black step function with shading
  indicating 95\% confidence limits) compared to the simulated CDFs
  (thin green band) generated from 100 SMOHALOS realizations.  The
  Kaplan-Meier method is applied to handle upper and lower limits.
  The total number of observations ($n$) is listed on the top along
  with the average difference in column density between observations
  and simulations ($\delta {\rm log} N$).  The three silicon ions
  ($\SiII$, $\SiIII$, \& $\SiIV$) are shown on top and $\CII$,
  $\MgII$, and $\OVI$ are shown on the bottom.  The input datapoints
  are shown along the bottom: squares for detections, upside-down
  triangles are upper limits, and upwards pointing triangles are lower
  limits.  Two-sided censoring results in the observed CDFs never
  reaching $0$ or $1$.  Vertical dotted lines encompass the range over
  which K-M estimation is not required.}
\label{fig:KM_all}
\end{figure*}

Figure \ref{fig:KM_all} shows the K-M estimator for six ions in
COS-Halos, which plots the cumulative distribution function (CDF) of
the fraction of absorbers with a higher column density in black with
shading indicating 95\% confidence intervals.  We apply a two-sided
Kaplan-Meier estimator to account for upper limits
(i.e. non-detections) and lower limits (i.e. saturated absorption) on
the observed column densities.  Vertical dotted lines in each panel
encompass the range over which the K-M estimator is not required and
uncensored detections set the CDF.  The K-M estimator applies to the
column density ranges where censored and uncensored data overlap, and
the K-M method uses the assumption that the censored data
distributions follow the uncensored data distributions.  This allows
us to use censored data points that have different detection limits to
estimate the probability distribution according to the detected
datapoints.  At the column density below (above) which all
observations become upper (lower) limits, the K-M estimator cannot
provide a constraint, therefore this assures that the CDF never
reaches 1 or 0 for all low ions, because these data are censored on
both sides.  

The main motivation of the two-sided K-M estimator is to achieve a
more statistically appropriate {\it and limited} CDF to compare with
simulated datasets.  For example, we limit the CDF at column densities
above which absorbers are all saturated, while an analysis like W13
assumes lower limits are uncensored detections, making a statistical
comparison with simulations more constrained.  Our application remains
agnostic about the true distribution of these lower limits, because it
is improper to assume saturated absorbers are at that column density
and in fact could be much higher as our simulations predict.  This is
critical for the interpretation of the COS-Halos data, because the ion
mass estimates would be under-estimated in such a case, which is a
point we further detail in \S\ref{sec:massest}.

To calculate the two-sided K-M estimator, we apply a normal one-sided
K-M estimator using the right-censored data (saturated lower limits)
and temporarily setting upper limits as detections.  We then apply
another one-sided K-M estimator using left-censored data (undetected
upper limits) setting the lower limits as detections.  The first K-M
estimator equals one minus the second K-M estimator between the
highest upper limit and lowest lower limit (i.e. between the vertical
dotted lines).  The two-sided K-M estimator uses the first K-M
estimator above the lowest lower limit and one minus the second K-M
estimator below the highest upper limit.  Our two-sided K-M estimator
relies on the highest upper limit always being lower than the lowest
lower limit.  If there is such a violation for a censored measurement,
we do not include it in our CDF.  However, this only happens for one
upper limit in $\SiII$ and $\CII$ from a very low signal-to-noise
(S/N) sight line and so we do not worry about the statistical bias.
It is thus rarely expected that upper limit non-detections overlap
lower limit saturated absorbers.

The requirements of the K-M method applied on a dataset as discussed
for absorption line measurements in \citet{sim04} are 1) the censored
datapoints must be independent of one another, and 2) the probability
that the datapoint will be censored should not correlate with its
value.  The first requirement is true, because the observed datapoints
for a given ion species come from different galaxies with no relation.
The second requirement is not true as upper limits depend on the S/N
of the spectrum, which is different between sight lines.  While this
second requirement is not fulfilled, we argue we can still apply the
K-M estimator owing to the implicit assumption that the observed
column densities have a much larger range of true uncensored values
compared to the range over which detections and censored limits are
observed.  Hence, we are arguing that the significant dispersion of
low ion columns, as predicted by the simulations
(cf. Fig. \ref{fig:b_SMOHALOS_prob}), makes their appearance as
censored datapoints in the COS-Halos sample essentially random.  This
is not true for $\OVI$ around blue galaxies, where the dispersion is
smaller than the observed range, but fortunately these datapoints are
almost all uncensored detections.  

The $\SiIII$ panel in Fig. \ref{fig:KM_all} demonstrates our K-M
estimator.  Upper limits overlap detections between
$N_{\SiIII}=10^{12.5}-10^{13} \cms$, and the K-M method applies the
probability distribution of the detections to the range of upper
limits where they overlap, as indicated by where the black line and
grey limits extend below the left dotted vertical line.  Below
$10^{12.5} \cms$, there are no detections, and the K-M estimator
provides no constraints.  Above $10^{13.2} \cms$, all $\SiIII$
absorbers are saturated, but there exist no detections to guide the
K-M prediction of non-detection and again there are no constraints on
the CDF.  This demonstrates the conservative nature of how we use the
two-sided K-M estimator, which limits us to comparing simulations and
observations where there exist detections.  There is more often
significant overlap between detections and upper limits than between
detections and saturated lower limits.  We urge caution when
interpreting the K-M estimator for column densities in the overlap
regions, owing to the assumption of censored data following uncensored
detections.  

\begin{figure*}
\includegraphics[width=0.98\textwidth]{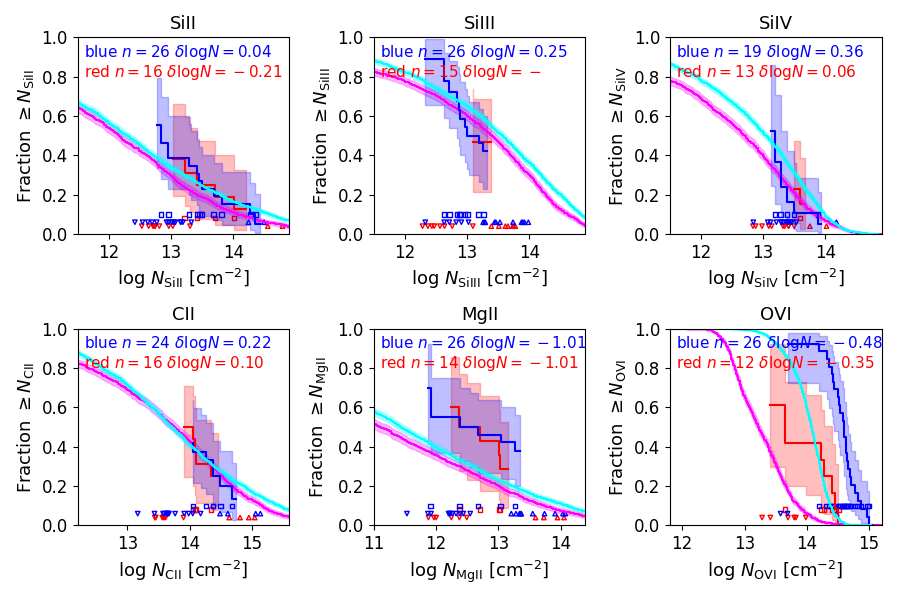}
\caption[]{CDFs using the Kaplan-Meier method of COS-Halos and
  SMOHALOS as in Fig. \ref{fig:KM_all}, but divided into blue and red
  galaxy samples.  COS-Halos observations are plotted in blue and red,
  and SMOHALOS simulations are plotted in cyan and magenta.  Sample
  statistics for blue and red galaxies are listed in each panel as in
  Fig. \ref{fig:KM_all}.  No $\delta {\rm log} N$ is given for the
  $\SiIII$ red sample owing to only having censored data points, but
  there still is a CDF measurement between the upper and lower limits.
  We do not plot the vertical dotted lines encompassing the ranges
  over which the K-M estimation is not required as we do in
  Fig. \ref{fig:KM_all}, but note that these ranges are equal to or
  larger for the individual blue and red samples as for the entire
  sample shown in that plot.}
\label{fig:KM_ssfr}
\end{figure*}

The K-M estimator lacks information on additional parameters beyond
the first, column density in our case, but we can sub-divide the
sample based on a second parameter, as we do for sSFR, and plot
multiple CDFs as in Fig. \ref{fig:KM_ssfr}.  While we lose information
on the dependence on impact parameter for each ion, we are generating
mock SMOHALOS surveys with the same impact parameters around similar
galaxies as observed.  This means that we do not generate mock
observations for COS-Halos galaxies when the given ion is not
observed, due to instrument coverage or blanketing by unrelated
absorbers.

\subsubsection{Results}

The K-M CDFs from 100 SMOHALOS realizations with perfect column
densities (i.e. no upper/lower limits) are shown in green in Fig.
\ref{fig:KM_all} and in cyan (blue sample) along with magenta (red
sample) in Fig. \ref{fig:KM_ssfr}.  Also listed for each
observation-simulation pair is the number of observations (uncensored
and censored, $n$) and the average column density deviation in dex of
the simulation from the observation ($\delta {\rm log} N$).  $\delta
{\rm log} N$ is calculated at each step in the observed K-M function
corresponding to each uncensored data point.  Observed data face
complete censorship on one or both ends of the column density
distribution, where the K-M functions cutoff before reaching 0 or 1.

The level of agreement between mock and real data in
Fig. \ref{fig:KM_all} varies from ion to ion.  The $\SiII$, $\SiIII$,
$\SiIV$, and $\CII$ total sample distributions usually agree within a
factor of $\approx 2$ or better ($0.3$ dex).  SMOHALOS $\MgII$ is
however $1.1$ dex too low according to this metric.  $\OVI$ is $0.5$
dex too low, which agrees with O16 and indicates that our slightly
modified simulated sample at $z=0.15-0.35$ is not different from O16's
sample.  The simulations overlap with the 95\% confidence limits of
the observations (large shaded bands) for the silicon and carbon
species, but not for $\MgII$ and $\OVI$.  $\CIII$ and $\CIV$ are not
shown due to their limited COS-Halos datasets, but the simulations
show reasonable agreement with COS-Halos.  $\CIII$ has 25
observations, of which only 2 are uncensored, while $\CIV$ only has 3
observations since COS-Halos was not designed to cover this ion.

Fig. \ref{fig:KM_ssfr} shows the subdivision into red and blue
galaxies.  As W13 showed, for the observed low ions the confidence
limits of the two galaxy samples always overlap.  Simulated red
galaxies have slightly lower column densities than blue galaxies with
the gap growing toward higher ions, but COS-Halos does not have a
sufficiently large sample to probe such small differences except for
$\OVI$.  We also sub-divide the blue sample into small and large
impact parameter bins, divided at 75 proper kpc (not shown), and the
results show statistically significant increases in column densities
at smaller impact parameters for all low ions, which agrees with W13.

Overall, the level of agreement between SMOHALOS and COS-Halos using
the K-M estimator is good for $\SiII$, $\SiIII$, $\SiIV$, and $\CII$,
being within a factor of two for the red and blue subsamples.  While
there exist some discrepancies-- simulated silicon ions are higher
than COS-Halos for blue galaxies and simulated $\SiIII$ has a larger
spread-- the simulations overlap the 95\% confidence limits for these
measurements.  While simulated $\SiII$ and $\CII$ column densities are
in reasonable agreement with observations, $\MgII$ is too weak for the
entire COS-Halos sample and any sub-division by sSFR and impact
parameter.  No self-shielding is included in these simulations, which
we next consider using standard equilibrium simulations.

\subsection{Model modifications} \label{sec:mods}

Our fiducial simulation results use non-equilibrium ionization
assuming a uniform \citet{haa01} background without self-shielding.
Therefore, we now explore NEQ ionization and the expected influence of
self-shielding on low ions.  We also comment on the effects of
simulation resolution (see also Appendix \ref{sec:res}) and other
sources of photo-ionization.

\begin{figure*}
\includegraphics[width=0.98\textwidth]{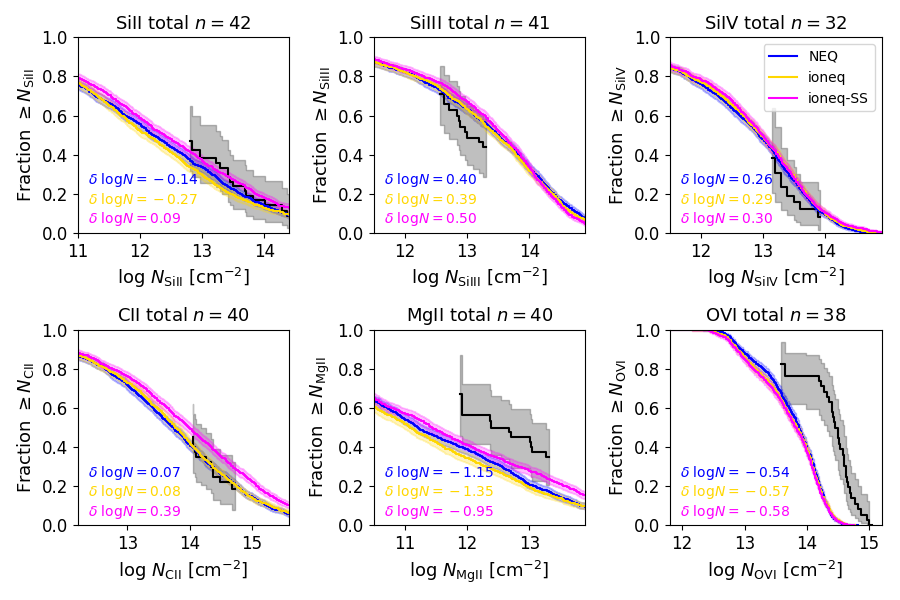}
\caption[]{Kaplan-Meier CDFs for the entire COS-Halos sample, as in
  Fig. \ref{fig:KM_all}, compared to different model variations.  The
  $z=0.20$ NEQ sample in blue is compared to standard equilibrium
  EAGLE simulations of the same haloes assuming ionization equilibrium
  for the uniform \citet{haa01} background (gold), and then applying a
  self-shielding criterion following the \citet{rah13} prescription
  (magenta).}
\label{fig:KM_SS}
\end{figure*}

Figure \ref{fig:KM_SS} compares the CDFs for our $z=0.20$ subsample
(NEQ in blue) to standard EAGLE equilibrium simulations where we
iterate $z=0.20$ outputs to ionization equilibrium using our NEQ
network for the same haloes (ioneq in gold), and then apply the
\citet{rah13} self-shielding correction (ioneq-SS in magenta).  To
simulate self-shielding in post-processing, we modify the regular NEQ
network by reducing the ionizing radiation for metal ions with
ionization potentials above $1$ Ryd according to the density and
redshift dependencies derived by \citet{rah13} from radiative transfer
simulations that reproduce the $\HI$ column density distribution.  The
simulation output is then iterated to this new self-shielded
ionization equilibrium.  This method is only an approximation because
it ignores the frequency dependent attenuation that declines for
higher ionization potentials.  However, multiply ionized species with
higher potentials are not appreciably photo-ionized at the densities
where the correction is used.  Thirty SMOHALOS realizations of each
model are run, and the baseline NEQ model shows essentially identical
behaviour as the full sample including all redshifts in
Fig. \ref{fig:KM_all}.

The NEQ and ioneq runs overlap for the most part, which we further
elaborate upon in Appendix \ref{sec:ioneq}-- NEQ ionization does not
significantly alter low ion abundances when assuming a uniform
ionization background.  $\MgII$ and $\SiII$ decline the most, but such
differences are expected given that the ioneq and NEQ simulations are
separate runs once the NEQ is turned on as described in
\S\ref{sec:runs}, and this does not mean there is a significant
difference that can be attributed to non-equilibrium ionization.

Applying the self-shielding criterion increases singly ionized species
at higher column densities as indicated by the average $\delta$log$N$
value increasing by $0.2-0.4$ dex for $\CII$, $\SiII$, and $\MgII$
over the ioneq model.  $\SiIII$ is also boosted by 0.1 dex.  This
slightly degrades the excellent agreement for $\CII$, yields a similar
fit for $\SiII$ as the NEQ model, and still leaves a factor of
ten-fold too small $\MgII$ column densities.  $\MgII$ traces the
highest densities of all these species, so it is not unexpected to see
the greatest increase due to self-shielding for higher $\MgII$ column
densities.

The under-prediction of $\MgII$ is concerning.  Part of this
discrepancy likely reflects that magnesium nucleosynthetic yields are
too low in EAGLE as \citet{seg16a} demonstrated that Mg in stars is
$\approx 0.3$ dex too low compared to other elements.  We also
consider the effect of resolution for a subset of $L^*$ haloes in
Appendix \ref{sec:res} and we find that $\MgII$ column densities
increase by $0.2$ dex when the mass resolution is increased by a
factor of eight over our standard runs, while $\CII$ and $\SiII$
decrease by $0.1$ dex.  This could result from these higher resolution
simulations better resolving dense sub-structure.  These effects
combined could raise $\MgII$ to overlap with the 95\% confidence
limits of the CDF in Fig. \ref{fig:KM_SS}.  However, they are unlikely
to simultaneously solve the $\OVI$ discrepancy.

The last model modifications we consider are additional sources of
photo-ionization from the central galaxy due to on-going
star-formation \citep[e.g.][W14]{sto13} and/or AGN.  The latter is
explored in \citet{seg17} and \citet{opp17} for EAGLE simulations such
as these where the addition of fluctuating AGN can enhance the
ionization levels even when the AGN is off as appears to be the case
for COS-Halos galaxies.  These works argue the proximity zone fossil
effect proposed by \citet{opp13b} is capable of enhancing $\OVI$
levels by $\approx 0.5$ dex for typical Seyfert-like AGN episodes in
star-forming galaxies.  The key is that the timescale to recombine
from higher ionization species to $\OVI$ is equal to or longer than
the typical times between AGN activity, even though the AGN is active
for only a small fraction of the time.  Proximity zone fossils can
solve the under-estimates of $\OVI$ in standard NEQ simulations (O16)
while not significantly reducing low ions, because even though low
ions are ionized to higher levels when the AGN is on, they rapidly
recombine to equilibrium after the AGN turns off \citep{opp17}.

The uniform ionization background may also be supplemented by a local
ionizing radiation field from emission sources within the galaxy
associated with ongoing star formation. Scaling this total local
ionizing flux with radius from the galaxy ($\propto r^{-2}$), galaxy
star formation rate ($\propto$ SFR), and the escape fraction of
ionizing photons ( $\propto$ $f_{\rm esc}$) was explored by W14 in
CLOUDY models.  The inclusion of stellar radiation from a Starburst99
spectrum \citep{lei99} can moderately affect COS-Halos results in the
inner CGM for fiducial values of star-forming galaxies in that survey:
SFR $=1 \msolaryr$ and an assumed $f_{\rm esc}=5\%$.  At the average
impact parameter in the COS-Halos blue sample, $b=72$ kpc, the
ionizing radiation from such a galaxy provides slightly more ionizing
radiation ($1.2\times$) than the \citet{haa01} background.  W14
consider these effects from the Starburst99 model and conclude that
this emission likely does not play a large role in setting the
ionization fractions of Si and O.  However, if we post-process our
simulation outputs using the physical densities predicted by the
model, then we see an increase of intermediate species like $\SiIII$
and $\SiIV$ and a decline in singly ionized species at $b\la 75$ kpc.
$\OVI$ remains mostly unchanged since it is collisionally ionized in
$L^*$ haloes at large radii, which appears to agree with \citet{sur17}
who found no difference outside 50 kpc while applying a more extreme
stellar radiation field that uses $f_{\rm esc}=5\%$ for lower energy
radiation and $100\%$ for soft X-rays.

On the other hand, not included in the Starburst99 models is the soft
X-ray emission produced by mechanical energy released into the ISM
during a starburst phase, from both supernovae and additional X-ray
sources produced by star-formation \citep{can10,wer16}, which may have
a substantial affect even at large impact parameters.  In contrast to
Starburst99 models that provide radiation mainly below $4$ Ryd, this
soft X-ray emission contributes only above $4$ Ryd and enhances $\OVI$
while not affecting low ions.
 
Our discussion of model modifications makes clear that there are
multiple potential effects that can alter low ion column densities by
factors of two or more.  Self-shielding can increase low ions while
extra ionization from star formation and AGN can reduce low ions.
Thus, the agreement of silicon and carbon species within a factor of
$\approx 2$ for the standard prescription can be classified as a
success of the model given these uncertainties.  The $\MgII$ column
densities are severely under-estimated, but could be remedied by going
to higher resolution, and by increasing the Mg yields, which show
evidence for being too low in EAGLE.

\section{Physical properties of low metal ions} \label{sec:phys}

Having explored how simulated observations compare to COS-Halos, we
now focus on the physical properties of the gas and metals traced by
low ions.  We first sum the metal mass budget traced by low ions and
follow up by linking observed ions to the physical properties of the
gas they trace.  The evolutionary state of CGM metals is considered
next.  Finally, we explore ion ratios used to constrain CLOUDY models,
e.g. in W14 and \citet{kee17}, and test the validity of single-phase
models.

\subsection{Low-ion CGM mass estimates} \label{sec:massest}

In O16, we used our zoom simulations to explore the circumgalactic
oxygen budget, finding that only $0.9-1.3\%$ of oxygen at $<R_{200}$
is in the $\OVI$ state for $L^*$ haloes spanning
$M_{200}=10^{11.8}-10^{12.2} \msolar$.  A much larger fraction of
oxygen inside the virial radius of those same halos, $27-52\%$,
resides in $\OI-\OIII$.  Having followed the NEQ ionization and
cooling in our zooms for 11 elements, we can self-consistently trace
the 15 silicon ion species in the same way as O16 traced the 9 oxygen
ion species.  $\SiII$, $\SiIII$, and $\SiIV$ comprise $19-42\%$ of
$L^*$ haloes' silicon budget ($\SiI$ is negligible), therefore these
``low''-ion silicon species provide a good proxy for the low-ion CGM
mass estimate.  Our simulations show that the $\SiI-\SiIV$ ion
fraction is consistently between $70$ and $80\%$ of the $\OI-\OIII$
ion fraction, which O16 plotted in their Figure 10.  The median
low-ion silicon CGM budget for $L^*$ haloes is
$3.9-4.5\times 10^6 \msolar$, which converts to
$7-8\times 10^7 \msolar$ for all metals using the simulation-averaged
Si/Z ratio.  Relative to solar abundances, our simulation-averaged
[Si/Z] and [Si/O] values are within $0.05$ dex of \citet{asp05}.

\begin{figure}
  \includegraphics[width=0.49\textwidth]{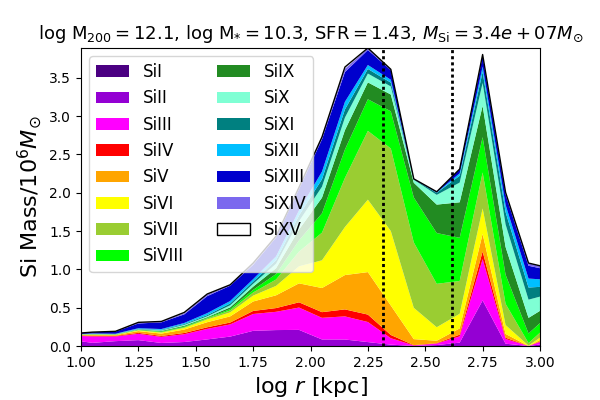}
  \caption[]{CGM silicon as a function of radius for a
    $M_{200}=10^{12.1} \msolar$ $z=0.2$ $L^*$ galaxy, subdivided by
    ion and summed in $0.1$ dex radial bins.  The total silicon
    mass budget is $3.4\times 10^7 \msolar$ between $10$ and $1000$
    kpc.  The dotted vertical lines indicate $R_{200}$ and $2R_{200}$,
    and the secondary bump beyond $2 R_{200}$ belongs to a
    neighbouring sub-$L^*$ galaxy. }
\label{fig:si_budget}
\end{figure}

Figure \ref{fig:si_budget} illustrates the breakdown of the silicon
budget around our reference $10^{12.1} \msolar$ $L^*$ halo at $z=0.2$
with shading indicating the contribution of various silicon ions as a
function of radius.  Purple, magenta, and red correspond to $\SiII$,
$\SiIII$, and $\SiIV$ respectively.  These low ions are primarily
found inside $R_{200}$ indicated by the left dotted line.  Significant
silicon at $T<10^5$ K exists also in $\SiV$ (orange) and $\SiVI$
(yellow).  Green and blue colours correspond to higher Si ions tracing
warm-hot CGM.  Most silicon (like all metals) resides beyond $0.5
R_{200}$, but low ions trace a biased set of interior metals.  

A second way to derive low-ion silicon masses corresponds to
integrating simulated, uncensored columns of $\SiII$, $\SiIII$, and
$\SiIV$ in the same $L^*$ haloes spanning $M_{200}=10^{11.8}-10^{12.2}
\msolar$ between $10$ and $150$ kpc.  This returns a low-ion silicon
mass of $5.2\times 10^6 \msolar$, corresponding to a total metal mass
of $9.6\times10^7 \msolar$.  These metals reside near the galaxy, with
just over half residing at impact parameters $10-25$ kpc, and only
15\% at $75-150$ kpc.  These two calculations are consistent although
slightly different, because the second one includes some ISM silicon
at impact parameters $\ga 10$ kpc, and the first one includes only the
CGM summed out to $R_{200}$, which is $\approx 200$ kpc rather than
$150$ kpc; the former appears to slightly outweigh the latter.

\begin{figure}
  \includegraphics[width=0.49\textwidth]{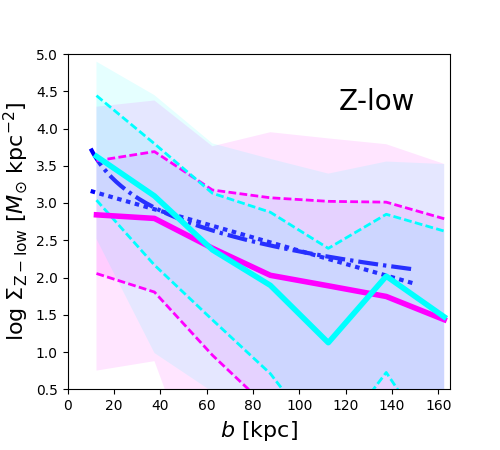}
  \caption[]{The median low ions metal surface densities for the 28
    star-forming galaxies (thick solid cyan line from 100 SMOHALOS
    realizations) derived from summing $\SiII$, $\SiIII$, and $\SiIV$
    compared to the COS-Halos low-ion metal surface densities derived
    from W14 CLOUDY modelling as reported in \citet{pee14} (blue
    dotted lines, 2 different functional fits shown).  One $\sigma$
    dispersions are indicated by dashed lines.  The passive SMOHALOS
    realizations are displayed in magenta for comparison.}
\label{fig:Zlowmass}
\end{figure}

We compare our values to those of \citet{pee14} derived from low-ion
CGM budgets traced by these silicon species and several other low
ions, finding an average metal mass of $2.3\times 10^7 \msolar$ within
150 kpc and $\delta v< 600 \kms$ of $L^*$ COS-Halos galaxies, but with
values up to $4\times$ higher, $\approx 9\times 10^7 \msolar$, when
including systematic uncertainties owing to ionization modelling.  We
compare the \citet{pee14} fits, derived from the ionization modelling
in W14, shown in dotted blue in Figure \ref{fig:Zlowmass} for the 28
star-forming COS-Halos galaxies.  Where W14 estimated low-ion metal
columns from CLOUDY modelling of uncensored and censored data, we sum
up $\SiII$, $\SiIII$, and $\SiIV$ SMOHALOS column densities, take the
mean as a function of $b$, and convert to a low-ion metal surface
density ($\msolar$ kpc$^{-2}$) assuming solar abundances.  The
comparison between the median SMOHALOS mass estimate (thick solid cyan
line) and the \citet{pee14} fits are promising, except for a dip in
the former at $75-125$ kpc.  Despite the SMOHALOS medians being at or
below the \citet{pee14} fits, we derive a $4\times$ higher low-ion
metal mass around $L^*$ galaxies owing to the significant dispersion
of column densities at a given impact parameter (dashed cyan lines
show the $1$-$\sigma$ dispersion).  The high-end \citet{pee14} mass
estimate of $\approx 9 \times 10^7 \msolar$ owes to their
consideration of systematic uncertainties in CLOUDY modelling, and not
the dispersion in column densities.

\begin{figure}
  \includegraphics[width=0.49\textwidth]{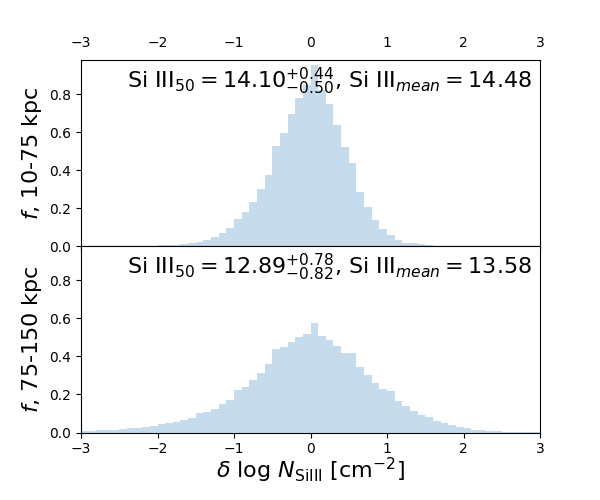}
  \caption[]{The residual distributions of $\SiIII$ column densities
    after subtracting the median column density profiles plotted in
    Fig. \ref{fig:bave_Nion}.  The distributions of $\delta
    N_{\SiIII}$ are calculated across 9 $L^*$ galaxies and binned into
    two impact parameter bins ($10-75$ kpc in the upper panel, and
    $75-150$ kpc in the lower panel).  For each bin, we list the
    median value of log $N_{\SiIII}$ and the $1-\sigma$ range, as well
    as the mean log $N_{\SiIII}$.  }
\label{fig:si3_disp}
\end{figure}

We emphasize the high metal mass value, $\approx 10^8 \msolar$,
indicating a significant reservoir of low ions, and argue that it is
consistent with COS-Halos.  The biggest difference between
\citet{pee14} and our summation is the treatment of the significant
scatter at a given impact parameter, which is also seen in COS-Halos
\citep[][their Fig. 7]{pee14}.  Figure \ref{fig:si3_disp} shows the
residual dispersion of $\SiIII$ among the $9$ $L^*$ halos, when
subtracting out the median $\SiIII$ binned in $\delta b=15$ kpc bins.
We divide the figure into two impact parameter ranges, $10-75$ and
$75-150$ kpc, to show that the residuals are well-described by a
log-normal distribution that increases in width at larger impact
parameters.  Using fewer galaxies or even just a single $L^*$ galaxy
results in similar dispersions.  The main point of
Fig. \ref{fig:si3_disp} is to show that low ion species are often
well-described by log-normal distributions at a given impact
parameter. We therefore suggest that ion mass estimates should
consider the median {\it and logarithmic dispersion} to sum up mass.
We list the median logarithmic column density ($\SiIII_{50}$) and
$1$-$\sigma$ range in the two panels, and also show that the mean
logarithmic column density ($\SiIII_{mean}$) is significantly higher
than the median.

We also wish to contrast our low-ion metal budget with the results of
\citet{mur16}, who found significantly fewer CGM metals in $\approx
10^{12}\msolar$ FIRE zoom simulated haloes.  They found between
$0.27-1.4\times10^8 \msolar$ of total metals in the CGM, while our
low-ion CGM component alone is $\approx 10^8 \msolar$.  \citet{mur16}
explained their lower CGM metal content has to do in part with FIRE
using lower yields than \citet{pee14}, the latter of which is similar
to our zooms (O16).  The \citet{mur16} zooms additionally have more
metals locked in stars (20-70\%) compared to our zooms (25-35\%, see
Fig. 9 of O16).  FIRE has not yet divided CGM metals into ionization
species, but we would predict that they would find smaller low ion
column densities than observed.  However the difference may not be as
large with newer FIRE zooms, since their recent m11.9a zoom has a
census more similar to ours, although it has a late-time merger that
recently enriched the CGM \citep{mur16}.

We also plot the galaxies with $M_{200} = 10^{12.7}-10^{13.2} \msolar$
in Figure \ref{fig:Zlowmass} and integrate a low-ion metal mass of
$6.6\times 10^7 \msolar$ using stringently isolated galaxies.  Thus
the low-ion content of group haloes is $\approx$2/3rd the amount of
$L^*$ galaxies within 150 kpc.  While we make the point that COS-Halos
passive galaxies likely have neighbouring galaxies that increase their
low ion column densities in \S\ref{sec:halotrends}, this isolated
group subsample nonetheless harbours a comparable amount of low-ion
metals within 150 kpc as is the case for $L^*$ haloes.

Finally, we tally the amount of cool CGM gas mass, defined as all
non-star-forming gas with $T<10^5$ K.  In $L^*$ haloes, the median
mass of cool gas is $1.5\times 10^{10} \msolar$ for a median halo mass
of $8.4\times 10^{11} \msolar$, and for group haloes, the cool gas
sums to $2.8\times 10^{10} \msolar$ for a median halo mass of
$7.2\times 10^{12} \msolar$.  We do not restrict to stringently
isolated galaxies for these sums, and there exists more cool gas
associated with satellites in group haloes compared to $L^*$ haloes.
We discuss mass budgets in future work, but note that the $L^*$ or
group sums are lower than the entire COS-Halos sample cool mass sum
within 160 kpc from \citet{pro17} of $(9.2\pm 4.3)\times10^{10}
\msolar$.


\subsection{Low-ion CGM physical properties} \label{sec:physprop}

\begin{figure}
  \includegraphics[width=0.48\textwidth]{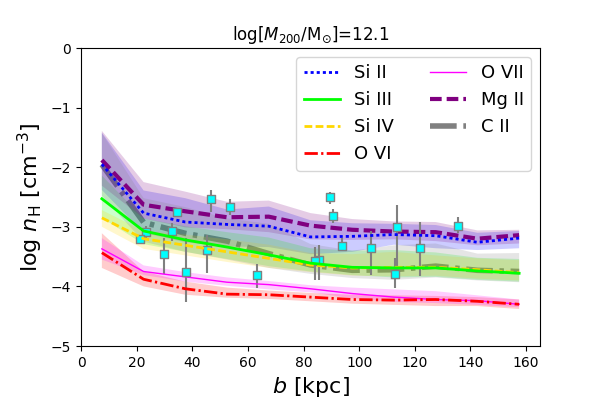}
  \includegraphics[width=0.48\textwidth]{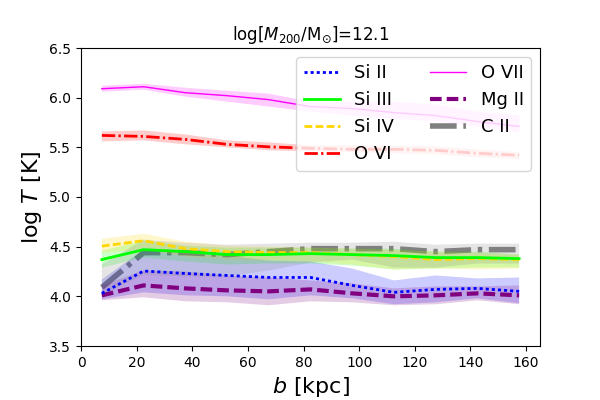}
  \includegraphics[width=0.48\textwidth]{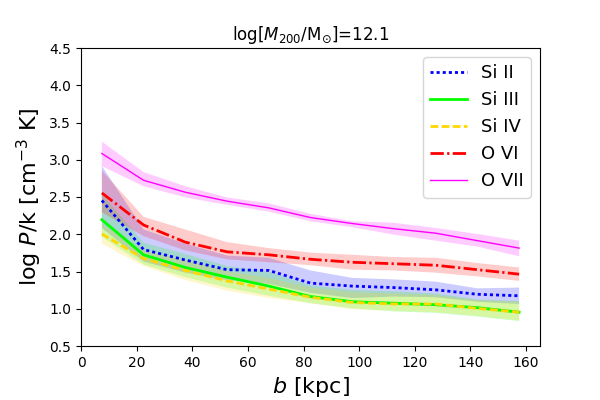}
  \caption[]{Physical properties weighted by ion column density as a
    function of impact parameter around the $10^{12.1} \msolar$ halo.
    Hydrogen number density and temperature are shown in the upper and
    middle panels for low species ($\SiII$, $\SiIII$, $\SiIV$, $\CII$,
    $\MgII$) as well as $\OVI$.  Additionally, $\OVII$ is shown to
    indicate the hot halo component, where most CGM metals reside in
    an $L^*$ halo.  Pressure, calculated by taking $\nh \times
    T/(X_{\rm H} \mu)$ is also plotted.  COS-Halos densities derived
    from low ion modelling \citep{pro17} around $L^*$ appear as cyan
    squares. }
\label{fig:b_phys}
\end{figure}

The column density ($N$)-weighted pixel value of a physical property,
$p$, is calculated according to
\begin{equation} \label{eqn:phys}
p_N = \frac{\sum\limits_{i} p_i \times N_i}{\sum\limits_{i} N_i}
\end{equation}
\noindent from column density maps where $i$ is a pixel with a column
density greater than $N_{\rm min}$.  We plot the median and
$1$-$\sigma$ spreads of $N$-weighted pixel values for density ($\nh$),
temperature ($T$), and pressure ($P$) as a function of impact
parameter in Figure \ref{fig:b_phys}.  We apply a minimum column
density based on typical observational column density limits:
$10^{12.5} \cms$ for Si species, $10^{13.5} \cms$ for $\OVI$, and
$10^{15.0} \cms$ for $\OVII$.  Weighted pixel values are not highly
sensitive to $N_{\rm min}$, although it prevents contributions from
pixels below observational capabilities.

The upper panel shows that the density that an ion traces declines
with ionization potential with little dependence on impact parameter
outside the inner CGM ($b\ga 50$ kpc) for our reference $10^{12.1}
\msolar$ halo.  Conversely, temperature increases with ionization
potential, while also showing little dependence on impact parameter:
low silicon ions clearly trace photo-ionized $T=10^{4-4.5}$ K gas,
$\OVI$ traces collisionally ionized warm-hot gas, and $\OVII$ traces
the $\approx 10^6$ K hot halo.  The trends of lower densities and
higher temperatures with increasing ionization potential and the weak
dependence on $b$ were also found in the \citet{for13} simulations and
form the basis of the \citet{ste16} universal density CGM model.
However, in contrast to our simulations, those models both have $\OVI$
photo-ionized in $L^*$ haloes.

Also shown for density and temperature relations are $\MgII$ and
$\CII$, where we use column density limits $10^{12.0}$ and
$10^{13.0} \cms$, respectively.  $\MgII$ absorbers trace denser and
cooler gas than $\SiII$, but still remain within $0.1-0.2$ dex of
$\SiII$ physical values.  $\CII$ traces gas more like $\SiII$ in the
very interior, but behaves more like $\SiIII$ and $\SiIV$ throughout
most of the CGM, because the ionization potential of $\CII$ ($24.4$
eV) is much higher than those of $\SiII$ ($16.4$ eV) and $\MgII$
($15.0$ eV).  We overplot the COS-Halos $\nh$ values derived modelling
these low ions \citep{pro17}\footnote{The \citet{pro17} densities are
  higher than the CLOUDY-derived densities of W14 (their Figs. 10 and
  12), owing to a factor of $4\times$ miscalculation in the latter,
  which did not account for the isotropic nature of the radiation
  field.}, which cluster around $\nh\sim 10^{-3} \cmc$ and show negligible
dependence with impact parameter, in reasonable agreement with the
simulation.

The lower panel of Fig. \ref{fig:b_phys} shows the pressure, computed
as $P/k=\nh T / (X_{\rm H} \mu)$, where $k$ is the Boltzmann constant,
$X_{\rm H}$ is the mass fraction of hydrogen, and $\mu$ is the mean
molecular weight.  The pressures traced by low silicon ions show a
mild decline with impact parameter for $b\ga 100$ kpc reaching
pressures $\sim 10$ $\cmc$K.  This is in agreement with the
\citet[][their Fig. 14]{sto13} CLOUDY-derived pressures determined for
their observed $z<0.2$ warm CGM clouds with impact parameters less
than $R_{\rm vir}$ around $L>0.1 L^*$ galaxies, which are comparable
to the star-forming COS-Halos galaxies.  The pressures derived from
W14 and \citet{pro17} are in a similar range as \citet{sto13}; hence,
our simulation predictions appear to agree with the densities and
pressures derived by CLOUDY modelling of COS low-ion metal column
densities.

\begin{figure}
\includegraphics[width=0.48\textwidth]{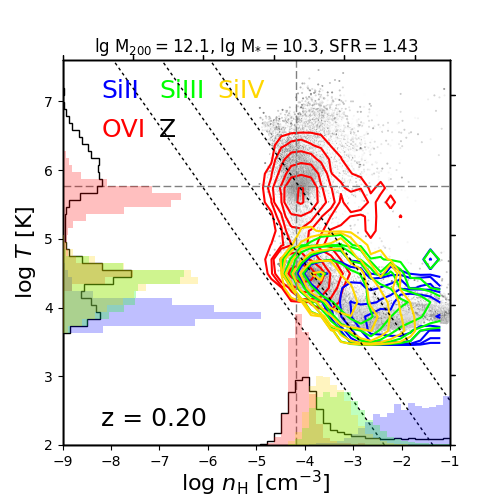}
\includegraphics[width=0.48\textwidth]{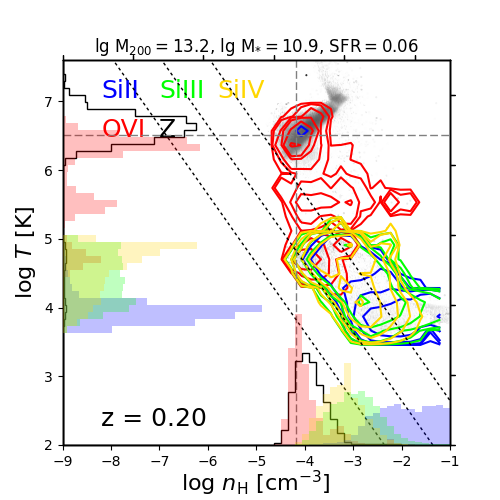}
\caption[]{Density-temperature phase space diagrams of $\SiII$ (blue),
  $\SiIII$ (green), $\SiIV$ (gold), $\OVI$ (red), and metals (grey
  shading) gas inside $R_{200}$ for a typical $L^*$ halo (upper panel)
  and group halo (lower panel) at $z=0.2$.  Corresponding histograms
  along the bottom and left indicate the density and temperature
  distribution for each ion, respectively. The dashed grey horizontal
  line indicates $T_{\rm vir}$ and the dashed grey vertical line
  indicates $200\times$ the critical overdensity.  Dotted diagonal
  lines indicate isobars at $1$, $10$, and $100$ $\cmc$K from left to
  right.}
\label{fig:rhoT}
\end{figure}

The density-temperature phase space diagrams in Figure \ref{fig:rhoT}
show the distribution of metals and ions inside $R_{200}$ for our
reference $L^*$ halo (upper panel) and a $M_{200}=10^{13.2} \msolar$
group halo (lower panel).  Dotted diagonal lines indicate isobars of
$P/k=1$, $10$, and $100$ $\cmc$K.  The two haloes show similar
distributions of low ions, $>10$ $\cmc$K, although the pressures are
slightly higher in the group-sized halo.  CGM metals peak at a density
$\nh = 10^{-4.1} \cmc$ for both halos, but the high-temperature
distribution peaks track the virial temperatures indicated by the
dashed horizontal lines, which increase according to $T_{\rm vir} \sim
M_{200}^{2/3}$ or a factor of $5.6\times$ across these two halos.

Applying a dividing line between hot and cool CGM metals of $T=10^5$
K, we find that about half the metals are in each phase in the $L^*$
halo, but only $3\%$ are in the cool phase for the group halo.
Surprisingly, there exists a similar mass of cool CGM metals inside
$R_{200}$ for the two of halo masses, $M_{\rm Z,cool} \sim
1.0-1.5\times 10^{8} \msolar$, which is borne out by observations of
low silicon species being similar between $L^*$ and group halos.  This
$M_{\rm Z,cool}$ value range holds among $L^*$ halos, but a $10^{11.8}
\msolar$ halo has as much as 75\% of its metals in the cool phase
while for a $10^{12.2} \msolar$ halo this fraction is as low as
$1/3$rd.  Therefore, the cool phase metal mass of a halo is nearly
invariant across a factor of over 20 in halo mass, while the total CGM
metal mass increases monotonically with halo mass.  Connecting with
the mass estimates in \S\ref{sec:massest} traced by low silicon ions,
of the order of half the cool, $T<10^5$ K CGM metals are traced by the
$\SiII-\SiIV$ phases with the rest in higher Si species tracing
densities $\nh\la 10^{-4.5} \cmc$.

\begin{figure}
  \includegraphics[width=0.48\textwidth]{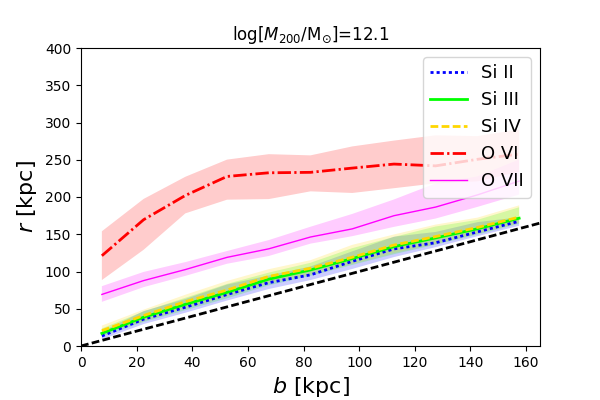}
  \includegraphics[width=0.48\textwidth]{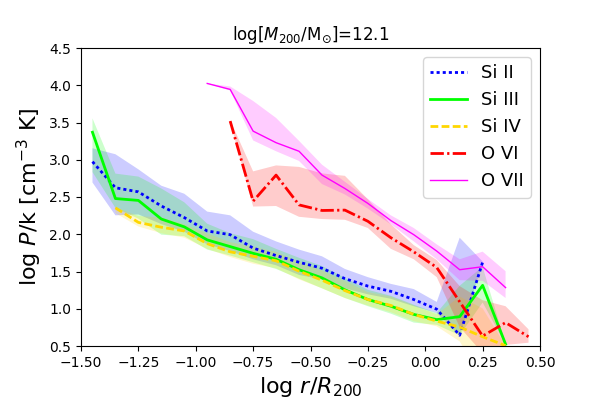}
  \caption[]{The upper panel shows column density-weighted mean
    physical radius, $r$, for a given ion plotted as a function of
    impact parameter for silicon species, $\OVI$, and $\OVII$ for the
    reference $M_{200}=10^{12.1} \msolar$ halo.  The $1$-to-$1$
    correspondence is plotted as a dashed black line.  Cool silicon
    species show $r \sim b$, while $\OVI$ in contrast arises from much
    larger physical radii at a given impact parameter (O16) and traces
    a physically and spatially distinct phase at a given impact
    parameter.  The lower panel plots ion-weighted pressure as a
    function of radius, indicating cool silicon ions and warm-hot
    oxygen species trace different pressures at the same physical
    radius.}
\label{fig:P_phys}
\end{figure}

Returning to radial trends, the upper panel of Figure \ref{fig:P_phys}
shows $N$-weighted radial distance as a function of impact parameter,
indicating that $\OVI$ arises at much larger physical radii than its
observed impact parameter.  This trend was noted by O16, who showed
that the typical $\OVI$ absorber observed at $b<150$ kpc traces gas at
$200-500$ kpc from the galaxy.  Conversely, $r_N \approx b$ for the
low silicon ions meaning that these ions are tracing gas at a physical
distance similar to their observed impact parameter.  One conclusion
is that the $\OVI$ phase is spatially distinct from low ions, so it is
not unexpected that $\OVI$ shows different kinematics from low ions
\citep{wer16}.  However, we note that \citet{wer16} found many aligned
components between low ions and $\OVI$, which may be a challenge for
our model here and is further explored in \citet{opp17} with the
inclusion of the proximity zone fossil effect.

Finally, we plot pressure as a function of radial distance in the
lower panel of Fig. \ref{fig:P_phys}.  The result is very different
from $P(b)$, because the $\OVI$ and $\OVII$ arise at much greater
radial distances than the observed impact parameters.  The low silicon
ions appear to be out of pressure equilibrium with the hotter phase
oxygen ions, although the ions are typically not spatially coincident.
The silicon ($\OVI$) ions are primarily at $r \la 0.5 R_{200}$ ($\ga
0.5 R_{200}$).  Nonetheless, in Appendix \ref{sec:press} we examine
the lack of pressure equilibrium at fixed radius, finding that some of
the low ion-traced clouds are embedded in a higher pressure ambient
medium, and discuss that this may be a numerical effect of the SPH
formalism indicating a lack of resolution as opposed to a physical
explanation.

Our result of low ions tracing $P\sim 10-100$ $\cmc$K gas while $\OVI$
also traces pressures at the higher end of this range is surprisingly
similar to the pressures calculated by \citet{mcq17} using their
analytic cooling flow model.  They argue that low ions are out of
pressure equilibrium with $\OVI$, indicating other non-thermal
pressure support for the low-ion clouds.  Our simulations find that
low ions are not spatially coincident with $\OVI$, but arise in the
interior CGM and are radially coincident with {\it even higher}
pressures traced by ions like $\OVII$ and above.  \citet{mcq17} argued
that the velocity alignment between low ions and $\OVI$ indicates they
are spatially related, suggesting low-ion clouds are directly cooling
out of the $10^{5.5}$ K phase.  We consider aligned absorbers in
\citet{opp17} where we argue that photo-ionized $\OVI$ in proximity
zone fossils, not included here, increases the amount of $\OVI$
components aligned with low metal ion components.  Without AGN
proximity zone fossils, there is still alignment between $\OVI$ and
low ions, although it is not as easy to line up individual components
between the two types of species.

Finally, we return to the factor of $>100\times$ higher cool CGM
densities predicted by the \citet{mal04} model versus the
observationally constrained results of W14.  The updated densities of
\citet{pro17} appear to cluster around $\nh \approx 10^{-3} \cmc$
reducing the tension a little bit, but still not close to the cool
phase densities predicted by W14 using the \citet{mal04} model of
$\sim 10^{-1.7}-10^{-1} \cmc$.  It is worth asking why there is so
much tension with this measurement given that the mean density of halo
gas should be of order $100\times$ the mean density of the Universe at
$z=0.2$, or $\nh \sim 10^{-4.5}\cmc$.  Using this mean density along
with the halo virial temperature of $10^{5.8}$ K for a $10^{12.1}
\msolar$ halo (O16), cool gas at $\sim 10^{4}$ K would require $\nh
\sim 10^{-2.7} \cmc$ for pressure equilibrium, which nearly matches
the \citet{pro17} results.  However, this implied pressure of $40$
$\cmc$K, while agreeing with low ion pressures in our simulations, is
much lower than the $\OVI$ and $\OVII$-traced gas in
Fig. \ref{fig:P_phys} at all radii except at $\ga R_{200}$.  The
disagreement with \citet{mal04} arises because their density and
temperature profiles steadily rise toward smaller radii where most
low-ion clouds are found.  Our density, temperature, and pressure
profiles also rise at smaller radii, which is why our low ion
pressures of $10-40$ $\cmc$K at $20-160$ kpc are out of pressure
equilibrium with the hot ambient medium at pressures $>100$ $\cmc$K.
There still exists tension between the COS-Halos-derived densities and
the densities predicted for pressure equilibrium with the inner, hot
CGM, but it is significantly less than the factor $\ga 100$ predicted
in W14.

\subsection{Low-ion CGM kinematic properties}  \label{sec:kinprop}

\begin{figure}
  \includegraphics[width=0.49\textwidth]{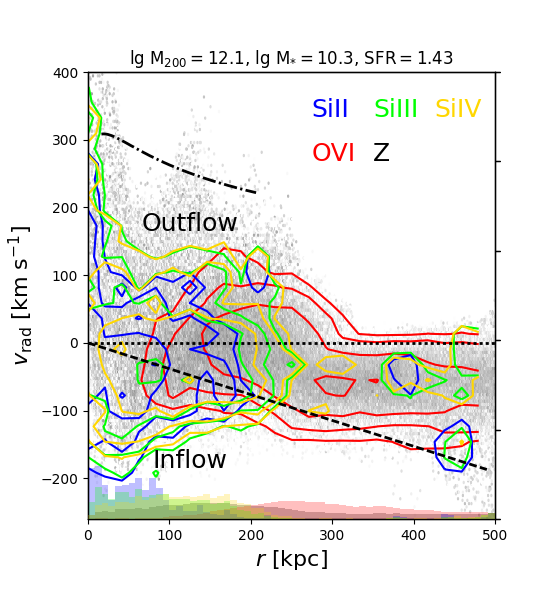}
\caption[]{Radial velocity relative to the central galaxy around the
  $M_{200} = 10^{12.1} \msolar$ reference halo for metals (grey
  shading) and various ions (coloured contours) as a function of
  radial distance.  Negative radial velocities indicate inward motion
  at $z=0.20$, and the dashed line delineates the velocity needed to
  reach the central galaxy by $z=0$.  The dot-dashed line indicates
  the escape velocity from the halo, showing very few winds are on
  trajectories that can escape the halo.  Coloured histograms
  correspond to the radial distribution of the ions.}
\label{fig:ion_rvel}
\end{figure}

We begin by considering the radial velocity of the gas particles
relative to the central galaxy, determined by calculating $v_{\rm rad}
\equiv \frac{v\cdot r}{r}$ as a function of radial distance in Figure
\ref{fig:ion_rvel} for our reference $M_{200}= 10^{12.1} \msolar$
halo.  Contours show that SPH particles with strong $\SiII$ are mainly
below the dashed line indicating the average inward velocity required
to reach the galaxy in the time elapsed between $z=0.20$ and $0$,
which is $2.5$ Gyr.  This indicates that most $\SiII$ has a velocity
trajectory consistent with gas accreting onto the galaxy by $z=0$.

SPH particles with strong $\SiIII$ and $\SiIV$ have radial
distributions weighted toward slightly larger radii than $\SiII$ (see
the bottom histograms), and most have negative, infalling velocities,
albeit slightly less than half of these Si ions have velocities
below the dashed line.  A small fraction of silicon species show
signatures of strong outflows ($v_{\rm rad}> 200 \kms$) in the inner
$\approx 30$ kpc, but even fewer exceed the escape velocity
from the halo (dot-dashed line) indicating that low ions rarely escape from
low redshift haloes.  $\OVI$, indicated by the red contours, has a
much larger radial extent and only a small fraction is on a velocity
trajectory that reaches the central galaxy by $z=0$, being mainly
above the dashed line.  Hence, the fate of the metals observed in the
CGM appears to be very different depending on its ionization state
\citep[e.g.][]{for14}.

It should be noted that even though the $\OVI$ is collisionally
ionized and shows a net inward flow, our simulations do not produce
the cooling flow structures theorized by \citet{hec02} and applied to
COS-Halos by \citet{bor16}.  Cooling behind a virial shock, at $\sim
R_{200}$ in that model, would produce a structured spread of
velocities inside the virial shock with $\OVI$ occupying a confined
post-shock region corresponding to efficient cooling around $T \sim
10^{5.5}$ K.  Instead, $\OVI$ occupies a range of radii, mainly
outside $R_{200}$ corresponding to $\sim 10^{5.5}$ K metal-enriched
gas with long cooling times, a large fraction of which also happens to
be inflowing.

\subsection{Evolution of CGM metals} \label{sec:metevol}

\begin{figure}
  \includegraphics[width=0.49\textwidth]{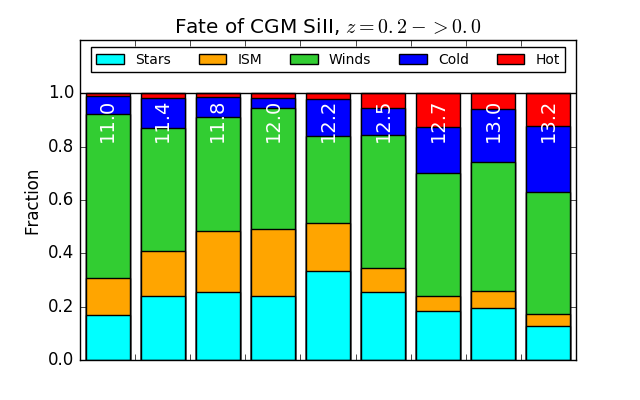}
  \includegraphics[width=0.49\textwidth]{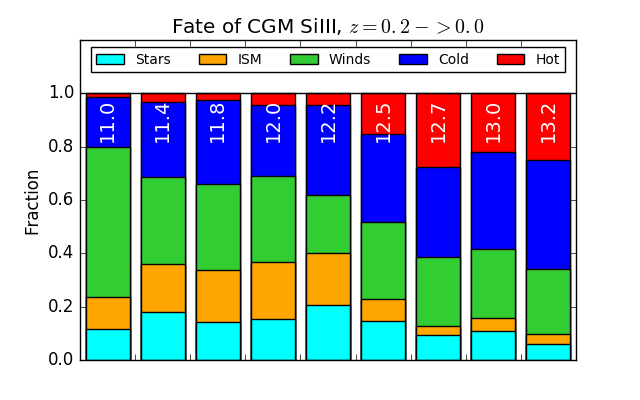}
  \includegraphics[width=0.49\textwidth]{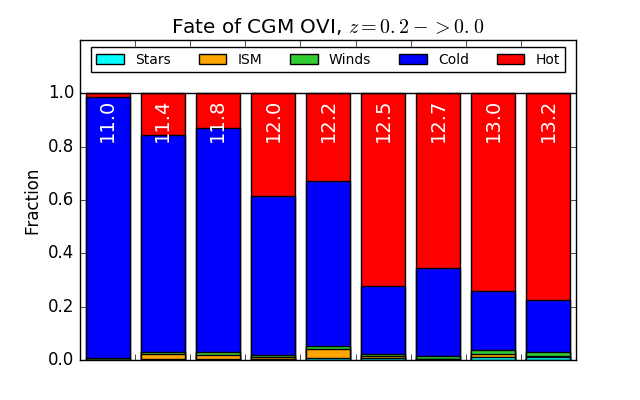}
\caption[]{Tracking of gaseous ion species ($\SiII$, $\SiIII$, and
  $\OVI$ from top to bottom) inside 500 kpc at $z=0.2$ for haloes
  increasing in mass from left to right (log[$M_{200}/\msolar$]
  indicated in white).  The bars indicate the fate of the gas traced
  by each ion at $z=0$, with the possibilities that it joined a galaxy
  (converted into stars, resides in ISM, or returned to the CGM) or
  remains in the CGM (cold and hot phases divided by $T=10^{5.0}$ K).
}
\label{fig:ion_futures}
\end{figure}

While the physical conditions of CGM gas can be ascertained by
observed metal ions, the evolutionary phase of the gas remains
difficult if not impossible to determine directly from observations.
Therefore, we use additional information within the simulations to
consider the $z=0$ fate of CGM metals observed at $z=0.20$.  Our
simulations allow particle tracking, so we can take subsets of gas and
determine if it reaches a galaxy or remains in the CGM.  We perform
this exercise on a range of halo masses and plot the phase fraction of
SPH particles above a specific ion threshold for a given ion species.
The fate of $z=0.2$ SPH particles for 9 haloes sorted by halo mass is
shown using the bar plots in Figure \ref{fig:ion_futures}.  Each bar
is divided into CGM SPH particles that 1) are converted into stars by
$z=0$, 2) are in the ISM at $z=0$, 3) have been recorded to be in a
galaxy's ISM between $z=0.2$ and $0$ (``winds''), but are in the CGM
and particles that have not been in the ISM and are 4) ``cold''
($T<10^5$ K) or 5) ``hot'' ($T\geq 10^5$ K) at $z=0$.  The threshold
ion fraction to select an SPH particle is 10\% $n_{\rm
  Si,\Zsolar}/n_{\rm H,\Zsolar} = 10^{-5.46}$ for silicon species and
5\% $n_{\rm O,\Zsolar}/n_{\rm H,\Zsolar} = 10^{-4.61}$ for $\OVI$--
this results in $\ga 10^4$ particles selected per $L^*$ halo and
corresponds to a level that creates significant absorption in a
spectrum.  We consider all particles within 500 kpc instead of
$R_{200}$ to include the extended $\OVI$ around $L^*$ haloes because
this barely affects silicon ions.  The 9 haloes plotted do not undergo
major mergers between $z=0.2$ and $0$, and because we do not
distinguish between accretion onto a central versus satellite, we
verified that most metals concentrate around the central galaxy.  The
exception is that satellite accretion dominates for $\SiII$ for haloes
with $M_{200} \geq 10^{13.0} \msolar$.

We show $\SiII$, $\SiIII$, and $\OVI$ in haloes ranging from sub-$L^*$
to our largest group ($M_{200}=10^{11.0}-10^{13.2} \msolar$).  Over
2.5 Gyr of evolution, more than half of every ion for every halo
resides in the $z=0$ CGM as defined by winds+cold+hot phases, but
there are large differences between species and trends with halo mass.
The vast majority of circumgalactic $\SiII$ observed at $z=0.2$ will
be accreted onto a galaxy by $z=0$, defined as stars+ISM+winds,
although more than half of this accretion will be re-ejected (winds)
into the CGM either by direct outflows (the majority) or by stripping
(a minority).  More of the $z=0.2$ circumgalactic $\SiII$ around
group-sized haloes remains in the CGM (cold+hot) indicating low
ionization metals are more extended and further from massive galaxies
relative to $L^*$ haloes (cf. Fig. \ref{fig:bave_Nion}), which results
in less accretion onto galaxies.  The wind component at $z=0$ is
primarily hot around all galaxies, because of the nature of our
thermal feedback.

In contrast, $\OVI$ rarely accretes onto any galaxy, which agrees with
the radii and radial velocities in Fig. \ref{fig:ion_rvel}.  Our focus
remains on the low ions, and $\SiIII$ shows behavior intermediate
between $\SiII$ and $\OVI$.  Most $\SiIII$ remains within 100 kpc of a
central galaxy for haloes $L^*$ and below, and most accretes onto the
galaxy.  Around group haloes, most $\SiIII$ and $\SiIV$ (not shown)
remains in the CGM, with the largest fraction remaining as cold CGM at
$z=0$.

\citet{for14} performed metal tracking in SPH simulations, finding the
same general trends we show here.  Like them, we find that lower
ionization state metals recycle onto galaxies more often, and that
lower mass haloes recycle more than higher mass haloes.  They also
showed that a low ion like $\MgII$ will fall onto a galaxy within
several Gyr in contrast to $\OVI$, which was injected by outflows many
Gyr ago (O16).  However, their $\OVI$ is primarily photo-ionized
\citep{for13} and a greater fraction of their $\OVI$ gas recycles back
onto their galaxies, in contrast with our primarily collisionally
ionized $\OVI$ that almost never recycles.  \citet{cra17} also
performed tracking of $\HI$ gas using EAGLE in their Figure 11, but
over a shorter timescale from $z=0.1\rightarrow 0$.  $\HI$ is most
comparable with our $\SiII$, and they found that most atomic hydrogen
remains associated with its galaxy (59\% remains $\HI$, 13\% turns
into stars) and a majority of the remaining 28\% is heated to high CGM
temperatures by feedback.

\subsection{Ion ratios} \label{sec:ionratios}

Ion ratios have long been used to derive physical properties of
absorbers.  For low ion absorbers that are photo-ionized, modelling
ion ratios can strongly constrain an absorber's ionization parameter,
which provides a measure of its physical density for a given
ionization field and temperature.  W14 used ion ratios from a variety
of carbon, silicon, nitrogen, magnesium, and oxygen species to derive
physical parameters of the metal-enriched CGM.  Their method involves
using ``integrated'' column densities where they sum the entire column
density along the sight line within $\pm 600 \kms$ of the galaxy
systematic velocity.  They found a single-phase CLOUDY model in thermal
equilibrium (but ignoring adiabatic cooling) by varying the ionization
parameter assuming the \citet{haa01} ionization background and
metallicity that matches the hydrogen column density of the sight
line.

\begin{figure}
  \includegraphics[width=0.43\textwidth]{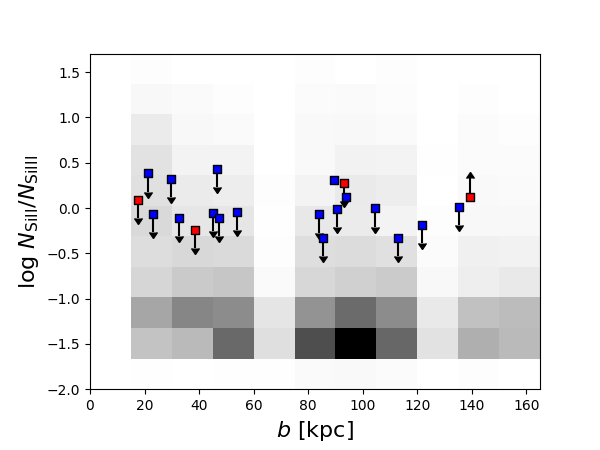}
  \includegraphics[width=0.43\textwidth]{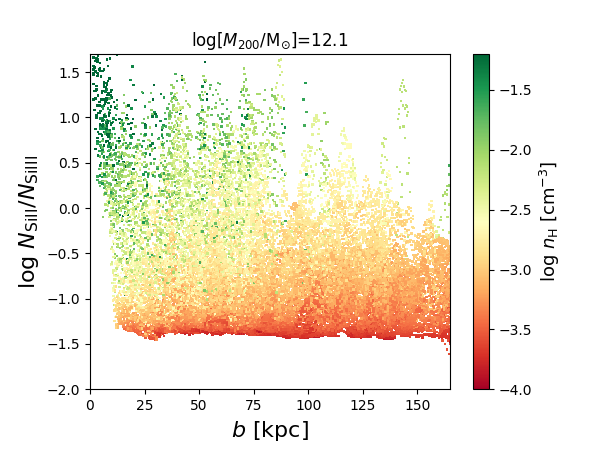}
  \includegraphics[width=0.43\textwidth]{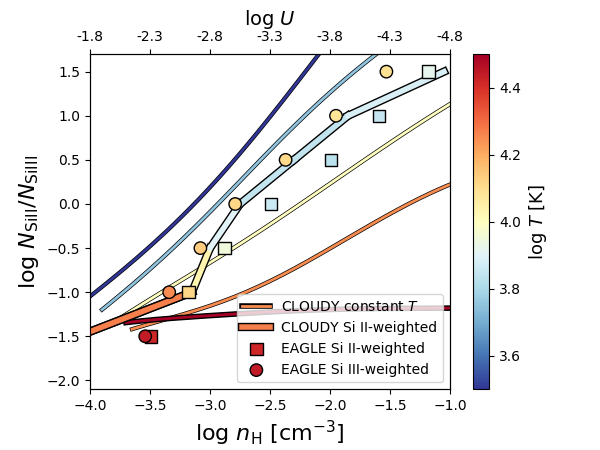}             
\caption[]{Ion ratio of $\SiII$ over $\SiIII$ as a function of impact
  parameter ($b$).  The upper panel shows COS-Halos blue and red
  subsamples from W13 with arrows indicating upper and lower limits.
  The SMOHALOS distributions (all ``detections'') are shown as gridded
  shading.  The middle panel shows the $\SiII / \SiIII$ ratio as a
  function of $b$ for all pixels where $N_{\SiII}$ and $N_{\SiIII}$
  are $>10^{12.5} \cms$, coloured by $\SiII$-weighted $\nh$ to show
  that this ratio is primarily sensitive to the physical gas density.
  The lower panel shows the relationship between
  $N_{\SiII}/N_{\SiIII}$ and $\nh$ in the simulations (data points),
  in CLOUDY models for various temperatures (thin lines), and
  variable-temperature CLOUDY models using $\SiII$-weighted $\nh$ and
  $T$ from the simulations (thick line).  Colours indicate the median
  temperature for each point.  The ion-weighted densities for the two
  ions ($\SiII$ \& $\SiIII$) differ, indicating that these ions probe
  a multiphase medium.  Fortunately, the close agreement between the
  thick line (CLOUDY models) and squares (simulated values) show that
  an density derived from an ionization ratio in a simulation gives a
  similar answer as a single-phase CLOUDY model, at least for
  $\SiII$-weighted values.}
\label{fig:Sifractions}
\end{figure}

Our fiducial simulations include NEQ effects, do not assume thermal
equilibrium, include adiabatic cooling, and unlike W14 do not include
self-shielding.  Self-shielding is important for higher densities,
especially above $\nh> 10^{-2} \cms$, but our analysis here focuses on
the $\SiII/\SiIII$ ratio, which mainly constrains lower densities.  We
show the COS-Halos $N_{\SiII}/N_{\SiIII}$ ratios, nearly all of which
are upper limits with $\SiII$ detected and $\SiIII$ a saturated lower
limit, in the upper panel of Figure \ref{fig:Sifractions}.  Data
points are coloured by $\SiII$-weighted $\nh$. SMOHALOS ratios with no
censoring are shown as grey gridded shading indicating mostly lower
$\SiII/\SiIII$ ratios consistent with the observed upper limits.  High
ratios are observed around two star-forming galaxies at 90 kpc and a
high lower limit is observed around a passive galaxy at 140 kpc, which
we argue next indicates dense gas.

To understand how the observed ion ratios relate to physical
parameters, we plot the integrated ion ratios for sight lines where
$N_{\SiII}$ and $N_{\SiIII}$ are greater than $10^{12.5} \cms$ as a
function of impact parameter for a projection of our reference $L^*$
halo in the middle panel of Fig. \ref{fig:Sifractions}.  There is a
clear relation between this ratio and $\nh$ that is mostly independent
of impact parameter.  High ratios are however seen preferentially in
the interior where there exists more dense clouds.  The lower envelope
of ratios owes to the precipitous decline in the $\SiII$ ionization
fraction at $\nh\la 10^{-3.7} \cmc$.  We choose to focus on the
$\SiII/\SiIII$ ratio because it is an adjacent ion ratio with high
sensitivity to $\nh$.

To compare the physical properties probed by the $\SiII/\SiIII$ ratio
in the simulation with those interpreted by the CLOUDY ionization
modelling as applied in W14, we show several relations between $\nh$
and $\SiII/\SiIII$ in the lower panel.  For each of our mock
integrated sight lines, we show two physical densities using
Eqn. \ref{eqn:phys}, a $\SiII$-weighted (squares) and a
$\SiIII$-weighted (circles) $\nh$.  We show the median densities for
several $\SiII/\SiIII$ bins to show that while they differ due to the
multiphase nature of the CGM, they remain within 0.5 dex of each
other, which matches the expected spread of densities probed by
$\SiII$ and $\SiIII$ in Fig. \ref{fig:b_phys} (blue and green lines in
the upper panel).  We colour data points by the median temperatures to
show that these vary as well.

CLOUDY modelling assumes a single-phase density and temperature, and
we plot the ratios inferred from CLOUDY models as thin solid lines for
5 different temperatures.  Although there is a significant dependence
on temperature as shown by the division between these lines, this
mainly owes to $\SiII$ becoming collisionally ionized above $\sim
10^{4.2}$ K and the ratio losing sensitivity to physical density.
Fortunately, when $\SiII$ is observationally detected, it is unlikely
to be collisionally ionized\footnote{Although low ions can be
  collisionally ionized, they almost always are photo-ionized in
  simulations and CLOUDY models, owing to rapid cooling at $T\ga 10^4$
  K to a thermal equilibrium at $T\la 10^4$ K at the high CGM
  densities where low ions are abundant.  We use $\SiII$-weighted
  quantities in the following analysis, because $\SiIII$-weighted
  quantities predict higher temperatures that would collisionally
  ionize $\SiII$ leading to the inability to use the $\SiII/\SiIII$
  ratio to predict density.} and we argue that the CLOUDY-derived
single phase model provides a reasonable constraint on $\nh$.  For an
integrated $\SiII/\SiIII$ ratio, the simulated $\SiII$-weighted $\nh$
agrees well with the CLOUDY-derived $\nh$ given that one selects the
correct temperature, which we show by the colour of the squares for
$\SiII$-weighted $\nh$.  To compare the squares to the CLOUDY models
of the correct temperature, we draw a variable-temperature thick line
that shows temperature-dependent CLOUDY models that use the
$\SiII$-weighted temperatures (colour of squares).  The good agreement
between the CLOUDY $\SiII$-weighted models and the EAGLE
$\SiII$-weighted densities instills confidence that ionization ratios
provide meaningful constraints on gas densities.  However, the
dependence on temperature is significant, and it is recommended in the
future that W14 and other similar works \citep[e.g.][]{kee17} publish
their assumed equilibrium temperatures derived from their ionization
parameters and metallicities.  The ionization ratio heavily depends on
this missing information.



The takeaway message is that CLOUDY ion ratio modelling of low ions
provides physical constraints that have the correct order of
magnitude, but the simulated CGM is a more complex multiphase medium
with adjacent low ions probing a spread in physical parameters.  Our
example here shows that the CLOUDY models best predict the physical
conditions traced by the lower ion, $\SiII$, which is why we show
$\SiII$-weighted $\nh$ in the middle panel of
Fig. \ref{fig:Sifractions}.  It is not unexpected that an ion ratio
such as $\SiII/\SiIII$ probes a multiphase spread in gas properties
given the different ionization potentials to ionize to these states
($8.2$ eV for $\SiII$, $16.3$ eV for $\SiIII$), but our exploration of
simulated physical properties provides confidence that CLOUDY
single-phase modelling yields physically meaningful constraints.
Therefore, it is not surprising that our simulations that provide good
matches to $\SiII$, $\SiIII$, and their ratios find similar densities
for low-ion clouds as \citet{pro17}.

We recommend expanding this type of analysis in future work.
Generating mock multi-ion observations from simulations, assessing
their goodness of fit, and determining the underlying physical
parameters of the simulation ion-by-ion is necessary to understand how
single-phase models used to constrain physical properties perform.  A
potential next step would be to perform this analysis on mock spectra
with individual components evaluated \citep[e.g.][]{wer16} instead of
an integrated system column density.

\section{Summary} \label{sec:summary}

We explore the low metal ions observed by the COS-Halos survey
\citep{wer13, wer14} using the set of EAGLE non-equilibrium (NEQ) zoom
simulations that \citet{opp16} found to reproduce the observed $\OVI$
bimodality around star-forming and passive galaxies.  Our exploration
considers $\CII$, $\CIII$, $\CIV$, $\MgII$, $\SiII$, $\SiIII$, and
$\SiIV$, with the primary focus on the silicon ions.  Simulated column
densities of silicon and carbon almost always agree with COS-Halos to
within a factor of two, although $\MgII$ is nearly a factor of ten too
weak.  Our main results regarding $L^*$ haloes
($M_{200}=10^{11.7}-10^{12.3} \msolar$) hosting star-forming galaxies,
and group-sized haloes ($M_{200}=10^{12.7}-10^{13.3} \msolar$) hosting
mostly passive galaxies are as follows:

\begin{itemize} 

\item{Simulated low metal ion column densities show 1) little
  dependence on galaxy sSFR, 2) a patchy covering fraction, and 3) a
  declining covering fraction at larger impact parameters.  Low ions
  trace a phase that is physically and spatially distinct from that
  traced by $\OVI$.  While low ions at observed impact parameters
  trace $10^4$ K clouds mainly within 100 kpc of both star-forming and
  passive galaxies, $\OVI$ traces the ambient $\sim 10^{5.5}$ K medium
  at radii $\ga 150$ kpc around $L^*$
  galaxies. (\S\ref{sec:halotrends}, Figs. \ref{fig:halotrends},
  \ref{fig:bave_Nion})}

\item{Simulated group galaxies often have neighbouring galaxies with
  $M_*>10^{10} \msolar$ inside 300 kpc, which is also true for some
  COS-Halos passive galaxies.  Neighbouring galaxies in and around
  group haloes increase the average low ion column densities within
  150 kpc.  However, when stringently restricting the sample to
  galaxies without neighbouring galaxies within 300 kpc, group haloes
  still show more low ion metals in the outer CGM at $> 100$ kpc than
  $L^*$ galaxies do. (\S\ref{sec:neighbours},
  Fig. \ref{fig:mhalo_Nave})}

\item{Going from the lowest ionization species ($\MgII$, $\SiII$,
  $\CII$) through intermediate species ($\SiIII$, $\SiIV$, $\CIV$) to
  $\OVI$, the dispersions and dependence on impact parameter of the
  column densities decline.  At the same time, the dependence on sSFR
  increases with $\OVI$ showing the strongest (i.e. $\OVI$-SSFR
  correlation, O16) and $\CIV$ showing the second strongest
  dependence. (\S\ref{sec:obs}, Fig. \ref{fig:b_SMOHALOS_prob})}

\item{Cumulative distribution functions of simulated silicon and
  carbon ion column densities overlap the 95\% confidence limits of
  COS-Halos with survival statistics applied to censored observations.
  This is true when considering the entire sample, sub-dividing by
  sSFR, and sub-dividing star-forming galaxies into two impact
  parameter bins.  The good agreement for C and Si is neither seen for
  $\MgII$ ($\approx 1$ dex too weak) nor for $\OVI$ ($\approx 0.5$ dex
  too weak).  (\S\ref{sec:surv}, Figs. \ref{fig:KM_all},
  \ref{fig:KM_ssfr})}

\item{Modifications to our fiducial NEQ simulation model can
  significantly alter column densities.  Self-shielding can raise low
  ion column densities by up to $0.4$ dex, while ionization from local
  sources can decrease low ion column densities but raise high ion
  column densities including $\OVI$.  The under-prediction of $\MgII$
  is likely in part caused by too low Mg yields in the EAGLE
  enrichment model. (\S\ref{sec:mods}, Fig. \ref{fig:KM_SS})}

\item{The total mass in metals traced by circumgalactic $\SiII-\SiIV$
  is nearly $10^8 \msolar$ inside $L^*$ haloes, with group haloes
  having $\approx 2/3$rd this amount.  The mass of cool ($T<10^5$ K)
  CGM metals inside $R_{200}$ is nearly invariant from $M_{200} =
  10^{11.8}-10^{13.2} \msolar$ at $1.0-1.5\times 10^8 \msolar$, while
  the fraction of CGM metals that are cool falls from 75\% to 3\% as
  haloes increase in mass and virial
  temperature. (\S\ref{sec:massest},\ref{sec:physprop},
  Figs. \ref{fig:si_budget}, \ref{fig:Zlowmass})}


\item{The pressures of low-ion metal clouds agree with \citet{wer14}
  and \citet{mcq17} with values of $10-40$ $\cmc$K for the CGM traced
  by COS-Halos.  Except near the galaxy, there is little dependence of
  $\nh$, $T$, and $P$ on impact parameter, which is also supported by
  \citet{sto13}, who found similar pressures independent of galaxy
  impact parameter.  The pressure of the $\OVI$ phase at $r \ga 150$
  kpc tends to be higher ($30-100$ $\cmc$K). (\S\ref{sec:physprop},
  Figs. \ref{fig:b_phys}, \ref{fig:rhoT}, \ref{fig:P_phys})}

\item{Most silicon ions observed in absorption at $z=0.2$ will accrete
  onto a galaxy by $z=0$, yet most of this accreted gas will be
  returned to the CGM mainly due to superwind feedback.  Higher
  ionization states and metals in higher mass haloes are less likely
  to accrete onto the galaxy over this 2.5 Gyr interval.  Silicon
  returned via superwind feedback mainly remains in a hot phase in the
  CGM. (\S\ref{sec:kinprop}, \ref{sec:metevol}, Figs. \ref{fig:ion_rvel},
  \ref{fig:ion_futures})}

\item{The silicon ion ratios observed by COS-Halos are broadly
  reproduced.  We show that single-phase CLOUDY models using ion
  ratios to calculate physical parameters yield densities of the
  correct order of magnitude.  The agreement is however imperfect
  because the gas is multiphase.  The $\SiII/\SiIII$ ratio is
  particularly constraining of physical density when adjacent low ions
  are observed. (\S\ref{sec:ionratios}, Fig. \ref{fig:Sifractions})}

\end{itemize}

The relatively good match between our simulations and COS-Halos is a
genuine prediction of the EAGLE model, because these simulations were
not calibrated to reproduce any CGM observations.  The low ions are
explored as a follow-up to the EAGLE $\OVI$ results showing the
bimodality of star-forming/passive galaxies in the CGM (O16), and
actually exhibit better agreement for silicon and carbon ion column
densities here than those $\OVI$ results that under-predict COS-Halos
column densities by about a factor of three.  Our results indicate
that low ions trace an almost completely distinct phase of the CGM
than $\OVI$-- one that traces the metal-enriched re-accretion of gas
onto galaxies.  It is therefore ironic that low ions do not reflect
the star-forming/passive bimodality of $\OVI$, which O16 argues is a
result of the $\OVI$ fraction peaking for the virial temperatures of
haloes hosting star-forming COS-Halos galaxies, while the passive
galaxies are predicted to reside in more massive haloes. This work
indicates that even low ions observed in $z\sim 0.2$ $10^{13}\msolar$
haloes will eventually accrete onto passive galaxies.

Combined with the results of O16 and from the AGN proximity zone
fossil mechanism \citep{opp13b, seg17}, we are building a unified CGM
model that can explain a wide range of observed metal column
densities.  The AGN proximity zone fossil mechanism, where AGN are
active for only a fraction of the time, can enhance $\OVI$ by the
necessary $\approx 0.5$ dex while not significantly altering low metal
ions \citep{opp17}.  This mechanism relies on most COS-Halos
star-forming galaxies being dormant or obscured AGN, which is argued
to be feasible.  However, radiation from local star formation could
also enhance $\OVI$ while also fitting the surprisingly weak $\NV$
column densities \citep{wer16}.  An important next step for testing
our unified CGM model will include the generation of mock spectra and
comparison of the kinematics of the low and high ions.

It is concerning that our good match for low ions relies on the
low-ion clouds being somewhat out of pressure equilibrium with the
higher pressure, hot medium.  Although this confirms other
calculations \citep[e.g.][]{mcq17}, our result likely indicates a
numerical issue with our implementation of SPH and insufficient
resolution at the phase transition (see Appendix \S\ref{sec:press}).
However, we stress that there is not a $>2$ order of magnitude
pressure imbalance implied in \citet{wer14} using the model of
\citet{mal04}.  New COS-Halos cloud densities are derived to be higher
\citep{pro17}, while EAGLE feedback reduces densities in the CGM
probed by COS-Halos by heating and transporting baryons to the outer
CGM, often beyond $R_{200}$.  Therefore, the pressure imbalance
between the cool clouds and the hot halo medium may be as little as a
factor of a few instead of greater than $100$.

Finally, we must also consider the under-prediction of $\MgII$,
especially when compared to the similar ions of $\CII$ and $\SiII$,
which match or slightly over-predict COS-Halos.  We identified several
modifications that could enhance this ion including self-shielding and
more realistic Mg yields than used in EAGLE, but higher resolution may
ultimately be needed to resolve the dense sub-structures that $\MgII$
traces.  Recent work by \citet{mcc16} suggested dense clouds below the
parsec scale could be responsible for dense CGM structures, which is
far beyond the capabilities of our simulations even though they show
good resolution convergence.  Depletion onto dust, which is also not
included in our simulations, could also alter the relative abundances.
Although, there is still a long way to go before we will fully
understand the complex, multiphase CGM observed by COS, we have shown
that simulations like ours can already guide the interpretation of the
observations and confront fundamental questions of how galaxies and
their haloes evolve across a Hubble time.

\section*{acknowledgments}

The authors would like to thank Joe Burchett, Ryan Horton, Ali
Rahmati, John Stocke, Todd Thompson, Colin Norman, Ryan O'Leary, and
Jason Prochaska for interesting conversations that contributed to this
manuscript. Support for Oppenheimer was provided through the NASA ATP
grant NNX16AB31G.  This work was supported by the Netherlands
Organisation for Scientific Research (NWO) through VICI grant
639.043.409.  This work used the DiRAC Data Centric system at Durham
University, operated by the Institute for Computational Cosmology on
behalf of the STFC DiRAC HPC Facility (www.dirac.ac.uk). This
equipment was funded by BIS National E-infrastructure capital grant
ST/K00042X/1, STFC capital grants ST/H008519/1 and ST/K00087X/1, STFC
DiRAC Operations grant ST/K003267/1 and Durham University. DiRAC is
part of the National E-Infrastructure.  RAC is a Royal Society
University Research Fellow.  AJR is supported by the Lindheimer
Fellowship at Northwestern University.

\appendix

\section{Ionization equilibrium test} \label{sec:ioneq}

\begin{figure*}
\includegraphics[width=0.98\textwidth]{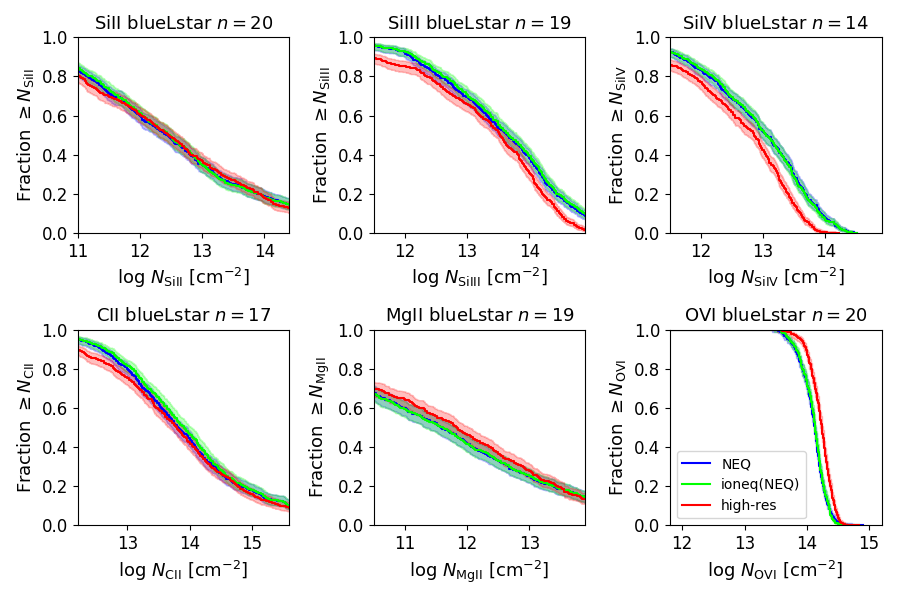}
\caption[]{SMOHALOS cumulative distribution functions simulating the
  COS-Halos blue $L^*$ galaxies using three zoom simulations hosting
  star-forming galaxies in haloes of $10^{12.1}-10^{12.3} \msolar$,
  all at $z=0.2$.  Blue indicates the standard NEQ simulations, green
  (often overlapping blue curves) indicates the NEQ simulations where
  ionization equilibrium is assumed, and red shows higher resolution
  NEQ {\it M4.4} resolution simulations.}
\label{fig:KM_res}
\end{figure*}

\begin{figure*}
  \includegraphics[width=0.325\textwidth]{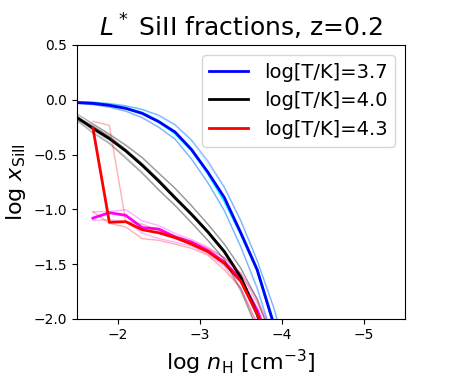}
  \includegraphics[width=0.325\textwidth]{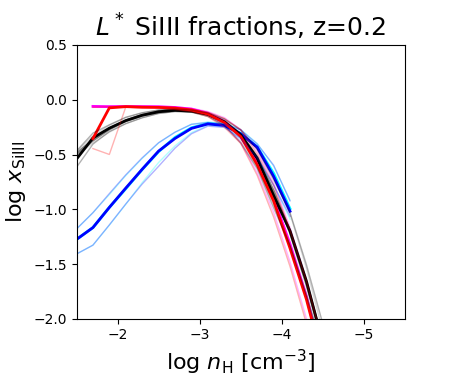}
  \includegraphics[width=0.325\textwidth]{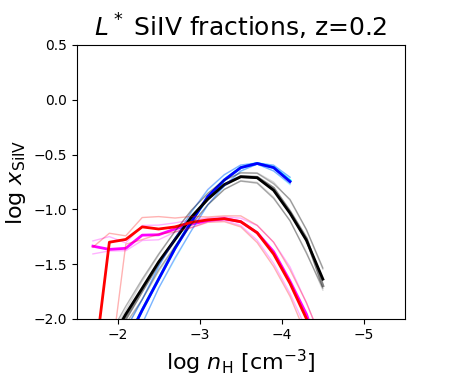}
  \caption[]{Median silicon ionization fractions, $x_{\rm ion}$, of
    gas particles as a function of density at $z=0.2$ for three
    temperatures ($T=10^{3.7}$ K in blue, $10^{4.0}$ K in black, and
    $10^{4.3}$ K in red) for the NEQ runs.  Particles are selected
    from $L^*$ zooms to lie within 300 kpc of the central galaxy.  We
    also show the ionization fractions for ioneq(NEQ) runs in cyan,
    grey, and magenta, and for the most part they lie directly beneath
    the NEQ curves except for instances where there is little data.
    The $1$-$\sigma$ spreads are also shown with thin lines and do not
    indicate a greater spread for NEQ versus ioneq(NEQ) runs.}
\label{fig:ionfractions}
\end{figure*}

Non-equilibrium effects are noted to be small when comparing
equilibrium and NEQ runs at $z=0.2$ in \S\ref{sec:mods}.  Figure
\ref{fig:KM_res} considers NEQ effects more in-depth by making a
different comparison: our standard NEQ runs (blue) are compared to NEQ
runs where we iterate our solver to ionization equilibrium in
post-processing (termed ioneq(NEQ) runs, green).  This differs from
the comparison than the one in Fig. \ref{fig:KM_SS} where the ioneq
runs are separate runs.  We use only three zoom simulations in haloes
of $10^{12.1-12.3} \msolar$ to simulate COS-Halos galaxies with
$M_*=10^{9.6}-10^{10.5} \msolar$ and sSFR$>10^{-11} {\rm yr}^{-1}$.
The comparison of the green and blue CDFs shows hardly any difference
between assuming ionization equilibrium and using the NEQ runs, which
bolsters our point that NEQ effects are generally unimportant for the
diagnostics of absorption line column densities.  To further
investigate NEQ effects, we show gas particle ionization fractions of
$\SiII$, $\SiIII$, and $\SiIV$ in Figure \ref{fig:ionfractions} as a
function of density selected from our $L^*$ zoom simulations.  Three
different temperatures ($T=10^{3.7}$, $10^{4.0}$, $10^{4.3}$ K) show
little difference between NEQ and the ioneq(NEQ) ionization fractions,
which overlap each other.

\section{Resolution test} \label{sec:res}

High-resolution {\it M4.4} simulations ($m_{\rm SPH} = 2.7\times 10^4
\msolar$) with $8\times$ better mass resolution than our standard runs
are shown in red in Fig. \ref{fig:KM_res}.  Their softening length is
175 proper pc below $z=2.8$, which is a factor of two smaller than our
fiducial resolution.  These runs show $0.1-0.3$ dex lower column
densities for silicon species and $\CII$ than standard {\it M5.3}
runs, but a $0.2$ dex higher column density for $\MgII$.  The {\it
  M4.4} runs show a similar increase in $\OVI$ of $\approx 0.1$ dex as
O16 showed in their exploration of {\it M4.4} runs compared to {\it
  M5.3} NEQ runs.

The {\it M4.4} runs have $0.18$ dex lower stellar masses than {\it
  M5.3} runs, which makes their stellar masses $3\times$ lower than
abundance matching constraints (O16), which is the main reason we do
not use these high resolution runs in the main portion of the paper.
The lower stellar masses are likely the result of improved numerical
efficiency of the stochastic thermal feedback and indicates $8\times$
higher resolution simulations need a recalibrated feedback
prescription as expected (see S15 for a discussion).  Because metal
production scales approximately with stellar mass, such a recalibration
may boost the column densities.  Interestingly, $\MgII$ moves $0.3$ dex
higher relative to $\CII$ and $\SiII$ when increasing to {\it M4.4}
resolution, which may help explain the $\MgII$ under-estimate.


\section{Pressures in SPH simulations} \label{sec:press}

\begin{figure*}
\includegraphics[width=0.48\textwidth]{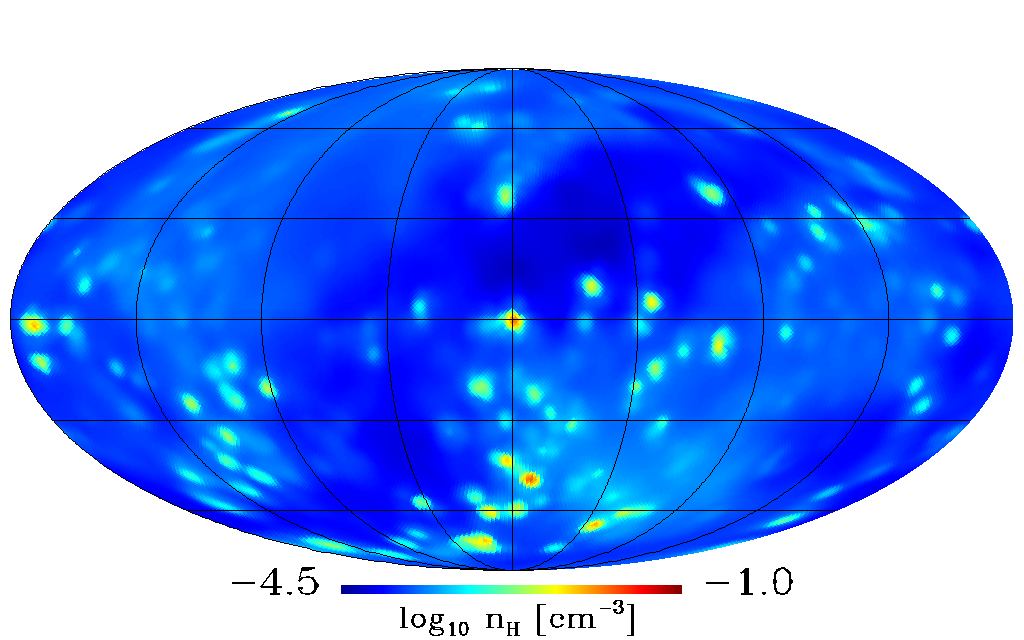}
\includegraphics[width=0.48\textwidth]{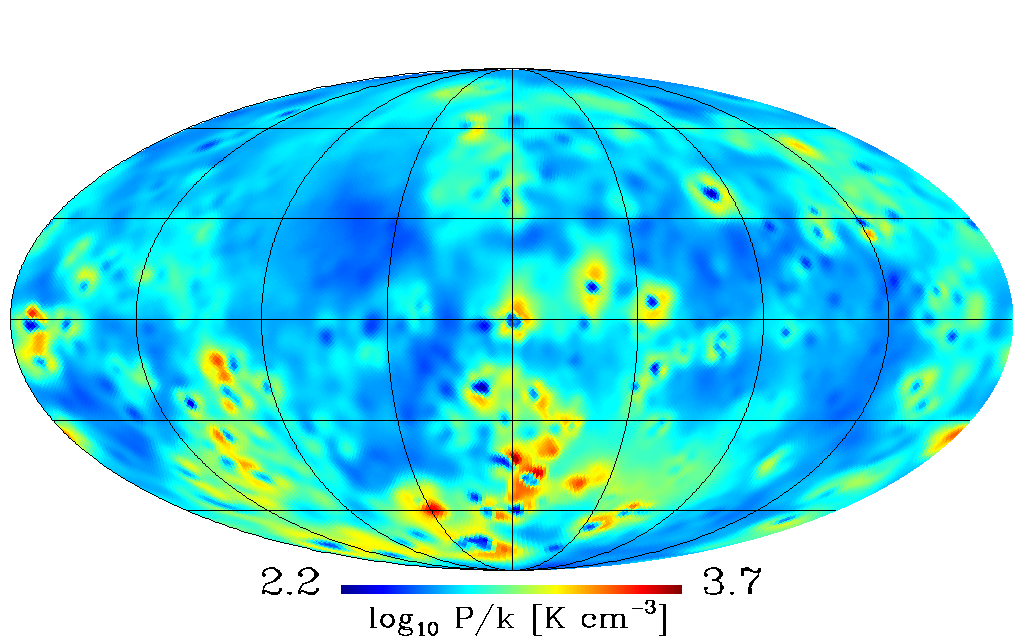}
\includegraphics[width=0.48\textwidth]{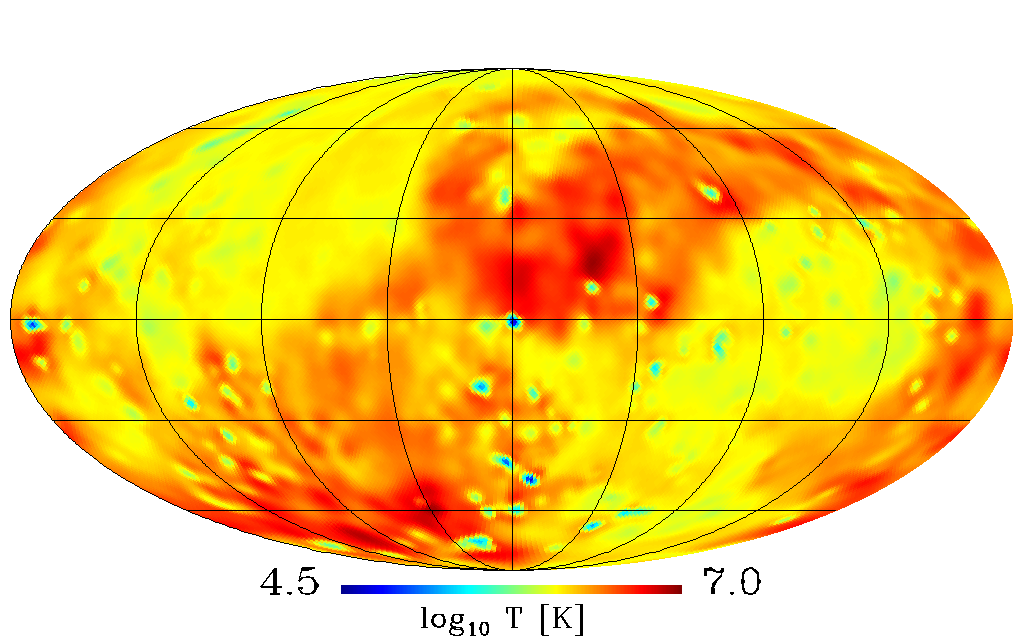}
\includegraphics[width=0.48\textwidth]{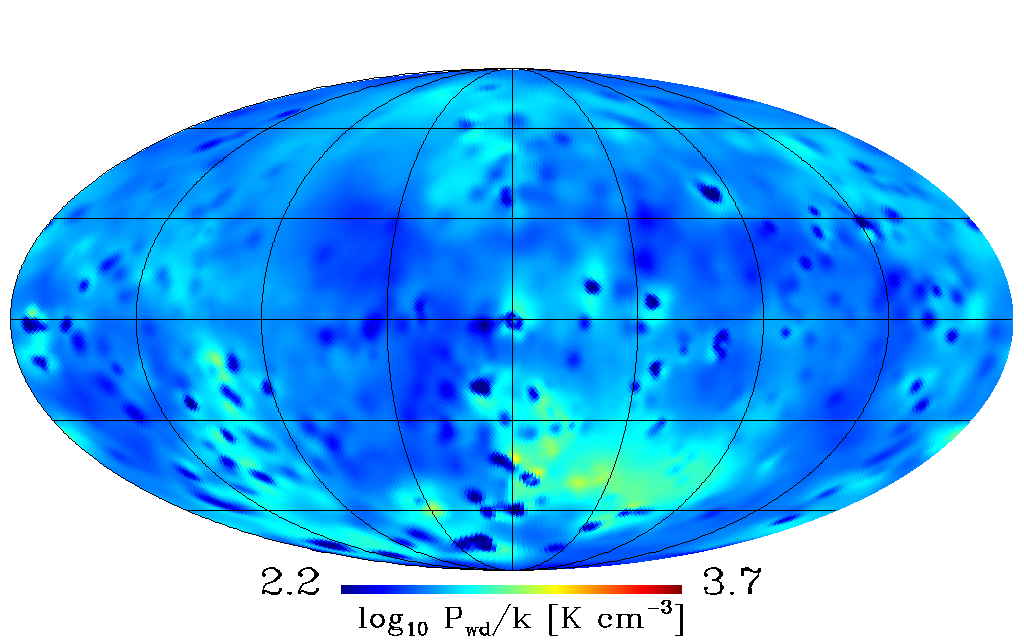}
  \caption[]{Mollweide projections of SPH particles at $0.3 R_{200}$
    of our reference $10^{12.1} \msolar$ halo.  Density and
    temperature (upper and lower left panels) show a multiphase
    structure of dense, cool CGM embedded in a hot ambient medium.
    The pressures (upper right panel) show much less variation between
    the phases, although some of the cool clumps appear to be at lower
    particle pressures.  The weighted pressure used in the hydro
    equations of motion is shown in the lower right panel and yields
    much smaller pressure differences.}
\label{fig:03rvirslice}
\end{figure*}

To explore why low-ion clouds appear to have lower pressures than the
hot, ambient medium, we consider whether these phases are spatially
coincident.  We display Mollweide projections of the CGM at $0.3
R_{200}$ of our reference zoom in Figure \ref{fig:03rvirslice}.  These
projections are made using the SPH smoothing lengths to project
particles onto a thin slice at $0.3 R_{200}$ (61 kpc) from the galaxy
center.  The two left panels show $\nh$ and $T$, indicating cool
$<10^5$ K clouds in an ambient $\ga 10^6$ K medium.  Clouds made up of
clumps of $T \la 10^5$ K and $\nh\ga 10^{-3}\cmc$ have a very small
filling factor in this kernel-smoothed slice projection.

In the upper right panel of Fig. \ref{fig:03rvirslice}, we plot $P/k$
by multiplying the particle $\nh/(X_{\rm H} \mu)$ and $T$, as
calculated in Figs. \ref{fig:b_phys} and \ref{fig:P_phys}.  Most cool
clumps appear as blue spots indicating lower pressures, albeit not as
low as the individual particle pressures owing to the smoothing
method.  There is variation in pressure across the projection, albeit
less variation than in $\nh$ and $T$.  However, some of the greatest
pressure differences occur around the cool clouds, which leads to a
discussion of the Anarchy formulation of SPH used in EAGLE
\citep[e.g.][]{schal15}.  The pressure-entropy SPH formulation
\citep{hopk13} aims to preserve pressure across a boundary.  However,
this pressure is calculated differently than our particle pressures as
it uses smoothed entropy and internal energy to determine the
``weighted'' pressure and the ``weighted'' density that affects the
hydrodynamic equations of motion.  Hence, the pressures we calculate
throughout this work are inconsistent with the pressure gradients
determining the dynamics.

When we calculate the weighted pressure, using a separately tracked
smoothed entropy variable, this pressure (lower right panel of
Fig. \ref{fig:03rvirslice}) displays much less variation around the
cold clumps.  For our exploration of low ions occupying the cold
clumps, ionization fractions likely do not differ much between using a
particle density versus a smoothed ``weighted'' density, $\nh_{\rm
  ,wd}$ as introduced in Equ. 8 of \citet{schal15} that relates to the
weighted pressure such that $P_{\rm wd}/k=\nh_{\rm ,wd} T / (X_{\rm H}
\mu)$.  The bigger difference would be in interface particles in
between the hot, ambient and cool, cloud phases, which will result in
different ionization fractions and cooling rates if we used the
weighted density.  The outer CGM harboring significant $\OVI$ at $r\ga
100$ kpc is not affected by these interface issues and represents a
predominantly single-phase CGM.

The difference between weighted and particle pressures at the cloud
interfaces indicated that the SPH scheme suffers from smoothing where
the calculation of cold cloud densities (particle and weighted) is
affected by lower density, ambient particles with smoothing lengths
overlapping even the centers of cold clouds.  This is fundamentally a
resolution issue, where smoothing lengths based on overlapping 58
neighbours using the C2 \citet{wen95} kernel will lead to a smoothed
gradient of physical parameters across a boundary.  Unfortunately,
while our clouds are resolved with multiple particles, their central
densities are affected by low-density ambient particles.  Higher
resolution could yield a different answer, but our resolution tests
here and in O16 show similar pressure differences between phases,
indicating the cloud masses scale with the SPH particle mass.  Hence,
the same issues persist at the {\it M4.4} resolution, but the clouds
are of lower mass.  Resolving the multiphase CGM spanning $2-3$ dex in
density and temperature at a boundary remains a challenging problem,
and the numerical behaviour of cold clouds in an ambient medium should
also be assessed using other hydro solvers.

\end{document}